\documentclass[11pt]{article}
\usepackage{amsmath,amsfonts,amsbsy}
\usepackage[dvips]{graphicx,epsfig}

 \csname
@addtoreset\endcsname{equation}{section}

\newcommand{\hoch}[1]{$^{#1}$}

\def\mso{\mathfrak{so}}
\def\msu{\mathfrak{su}}
\def\muu{\mathfrak{u}}

\def\msp{\mathfrak{sp}}

\def\mhs{\mathfrak{hs}}
\def\mhso{\mathfrak{ho}}
\def\mg{\mathfrak{g}}
\def\mh{\mathfrak{h}}

\def\Real{{\mathbb R}}
\def\Comp{{\mathbb C}}
\def\integ{{\mathbb Z}}
\def\id{{\mathbb I}}

\def\ncross{{}_{\times}\!\!\!\!^{\times}}  

\def\bec{\begin{center}}
\def\ec{\end{center}}
\def\a{\alpha}  

\def\b{\beta}

\def\d{\delta} 

\def\e{\epsilon}

\def\l{\lambda}

\def\s{\sigma}
\def\S{\Sigma}
\def\t{\tau}

\def\x{\xi}

\def\O{\Omega}

\def\sfD{{\mathfrak D}}
\def\sfV{{\mathfrak V}}
\def\sfN{{\mathfrak N}}
\def\sfDid{{\widehat{\mathfrak D}_{[0]}(0,0)}}
\def\sfDid{{\widehat{\mathfrak D}_{[0]}(0,0)}}
\def\sfDq{{\widehat{\mathfrak D}}_{[0]}(-\ft12,\ft12)}

\def\cN{{\cal N}}

\def\cS{{\cal S}}

\def\del{\partial}

\let\la=\label

\def\nn{\nonumber}
\newcommand{\eq}[1]{(\ref{#1})}

\def\be{\begin{equation}}
\def\ee{\end{equation}}
\def\bea{\begin{eqnarray}}
\def\eea{\end{eqnarray}}
\def\ba{\begin{array}}
\def\ea{\end{array}}

\def\mx#1#2#3#4{\left#1\begin{array}{#2} #3 \end{array}\right#4}

\def\ft#1#2{{\textstyle{{\scriptstyle #1}
\over {\scriptstyle #2}}}}

\def\scs#1{\section{\sc #1}}
\def\scss#1{\subsection{\sc #1}}

\begin{document}

\hfill{\texttt{SPIN-06/47}}

\vspace{-5pt}

\hfill{\texttt{ITP-UU-06/57}}

\vspace{-5pt}

\hfill{\texttt{KUL-TF-06/35}}

\vspace{-5pt}
\hfill{\texttt{hep-th/0701051}}

\hfill{\today}

\vspace{20pt}

\begin{center}


{\Large\sc On Singleton Composites in \\[10pt] Non-compact
WZW Models}


\vspace{30pt}
{\sc J. Engquist\hoch1, P. Sundell\hoch2 and L. Tamassia\hoch3}\\[15pt]

\hoch{1}{\it\small Institute for Theoretical Physics and Spinoza
Institute \\ Utrecht University \\3508 TD Utrecht, The
Netherlands}\vspace{5pt}

\vspace {0.5cm}

\hoch{2}{\it\small Scuola Normale Superiore and INFN\\ Piazza dei
Cavalieri 7, 56126 Pisa, Italy}\vspace{5pt}

\vspace {0.5cm}

\hoch{3}{\it\small Instituut voor Theoretische Fysica
\\Katholieke Universiteit Leuven \\
Celestijnenlaan 200D, B-3001 Leuven, Belgium}\vspace{30pt}


{\sc\large Abstract}\end{center}

\noindent We examine the $\widehat{\mso}(2,D-1)$ WZW model at the subcritical
level $-(D-3)/2$. It has a singular vacuum vector at Virasoro level $2$. Its
decoupling constitutes an affine extension of the equation of motion
of the $(D+1)$-dimensional conformal particle, \emph{i.e.}~the
scalar singleton. The admissible (spectrally flowed) representations
contain the singleton and its direct products, consisting of
massless and massive particles in $AdS_D$. In $D=4$ there exists an
extended model containing both scalar and spinor singletons of
$\msp(4)$. Its realization in terms of $4$
symplectic-real bosons contains the spinor-oscillator constructions
of the $4D$ singletons and their composites. We also comment on the
prospects of relating gauged versions of the models to the
phase-space quantization of partonic branes and higher-spin gauge
theory.

{\vfill\leftline{}\vfill \vskip  10pt \footnoterule {\footnotesize
E-mails: \texttt{J.Engquist@phys.uu.nl, p.sundell@sns.it,
laura.tamassia@fys.kuleuven.be} \vskip -12pt}}

\setcounter{page}{1}

\pagebreak

\tableofcontents


\scs{Introduction}\label{Sec:1}


At the core of Quantum Field Theory (QFT) lies the representation
theory of non-compact Lie algebras. Their non-trivial unitary
irreducible representations (UIR) are necessarily
infinite-dimensional, with energy bounded from either below or
above, corresponding, in the first-quantized language, to the
presence of particles and anti-particles. In this paper, we shall
examine this feature in the $\widehat{\mso}(2,D-1)$
Wess-Zumino-Witten (WZW) model with subcritical level $k=-(D-3)/2$,
with particular focus on the case of $D=4$.

Non-compact WZW models furnish a novel mathematical topic that
remains relatively uncharted in comparison with the compact case,
essentially due to the above-mentioned fact that
infinite-dimensional UIRs arise already at the level of the
finite-dimensional subalgebra. Previous investigations related to
QFT in anti-de Sitter (AdS) spacetime have considered cases with
\emph{non-critical} level $k+h^{\vee}<0$
\cite{Fradkin:1991mw,Bars:1989ph,Hwang:1998tr,Maldacena:2000hw},
\emph{critical} level $k+h^{\vee}=0$
\cite{Bakas:2004jq,Lindstrom:2003mg} as well as \emph{subcritical}
levels $k+h^{\vee}>0$ \cite{Dobrev:1990ng}, where $h^{\vee}$ is the
dual Coxeter number given by $h^{\vee}=D-1$ for $\mso(2,D-1)$.

The attempts of interpreting the critical case as a form of
tensionless limit have been problematic, technically, at the level
of two-dimensional conformal field theory (CFT), where the standard
Sugawara construction breaks down. In this respect, the subcritical
cases are more tractable. Interestingly, already in
\cite{Dobrev:1990ng} it was found that in $D=4$ the subcritical
$k'=-5/2$ Verma module built on the scalar singleton exhibits a
large number of singular vectors, signalling some form of symmetry
enhancement (one might also speculate about some form of duality
between subcritical models with levels $k$ and $k'$ obeying
$k+k'=D-1$). More recently, starting with the work in
\cite{Witten:2003nn}, attempts to formulate string theory duals of
$\cN=4$ Super-Yang-Mills theory have led to CFTs
\cite{Berkovits:2004hg,Siegel:2004dj,Berkovits:2004jj} that are
tantalizingly close to our subcritical cases, although the precise
relation remains to be seen. $\widehat{\mso}(2,D-1)$ WZW models in four dimensions have also been considered in the literature and, for the choice $D=5$, it has been proven that a dynamical sector that is Einstein's general relativity arises \cite{Anabalon:2006fj}.

Our physical interest in the subcritical models with $k=-(D-3)/2$
(in the case of purely bosonic models) stems from the fact that they
exhibit \emph{compositeness}, whereby physical particles, such as
photons and gravitons, instead of being fundamental, are made up
from more elementary constituents, known as \emph{singletons}.
Drawing on the group theoretical underpinning of the AdS/CFT
correspondence and the features of Vasiliev's unfolded approach to
Higher-Spin Gauge Theory (HSGT) and QFT in general (for a recent
review, see \cite{Bekaert:2005vh} and references therein; for a
recent development in the direction of M-theory see
\cite{West:2007mh}), one arrives at the idea that the singletons are
the basic constituents of String Field Theory (SFT) and that this
feature is manifest when SFT is expanded around tensionless AdS
backgrounds -- while being blurred in expansions around tensionful
backgrounds including flat spacetime.

A key step in entertaining this partonic picture is to find a
connection between the standard background-dependent quantization of
strings and a covariant formulation that combines phase-space
quantization of discretized tensionless strings with unfolded
dynamics. Roughly speaking, the worldsheet correlators -- which are
normally expanded around metric backgrounds with finite string
tension in terms of cut-off field-theoretic Feynman diagrams --
should be given an expansion around a topological vacuum with zero
tension, resulting in algebraic structures making up the internal
sector of unfolded SFT. In the unfolded approach, the total manifold
is a non-commutative fiber (the string phase space) times a
commutative base manifold (containing the actual physical spacetime),
and the unfolded field equation is the (non-linear) cohomology of
the phase-space BRST operator plus the exterior derivative. As it is
the case already in HSGT, the projection to the base then yields a
Free Differential Algebra (FDA) constituting a quasi-topological QFT
with infinitely many zero-forms. Here, spacetime, instead of being a
slice of the non-commutative phase space, reemerges, fully
covariantly, upon fixing the manifest homotopy invariance
of the FDA.

Put somewhat differently, we have in mind a worldsheet duality
between, on the one hand, the standard BRST formulation based on the
Virasoro algebra and sewing of surfaces, and, on the other hand, a
non-standard BRST formulation (yet to be spelled out in detail)
based on gauging subcritical affine algebras and sewing of string
partons. As we shall see in this paper, and which is of great
conceptual interest, the subcritical affine algebras incorporate the
key features of the corresponding, already known, higher-spin
enveloping algebras, thus, seemingly, superseding the latter as the
key algebraic structure underlying unfolded SFT.

In a previous paper, \cite{Engquist:2005yt}, two of the authors of
the present paper have examined ordinary bosonic branes in $AdS_D$
using four related models: \emph{i)} the normal-coordinate expansion
around Nambu-Goto solitons describing rotating branes; \emph{ii)}
conformal branes close to the boundary of the AdS spacetime;
\emph{iii)} discretized tensionless branes on Dirac's hypercone; and
\emph{iv)} a topological non-compact gauged subcritical WZW model
(closed singleton strings) in $D=7$. It was found that
\emph{(i)-(iii)} realize the \emph{brane partons as scalar
singletons}. The same was conjectured to hold also in the WZW model
\emph{(iv)}, and, as already announced, this seems indeed to be
the case, as we shall demonstrate in this paper (see Section
\ref{Sec:3}). Out of these models, we expect the WZW-model formulation
to be the most viable one, since it entails the incorporation of a
``stringy'' multi-parton spectrum -- with massless as well as
massive representations -- into the above-mentioned unfolded
approach to QFT. That is, as outlined above, the WZW model realizes
an algebra of functions on a phase space (given, roughly, by the
direct sum over direct products of singleton phase spaces) to be
identified as the fiber of an unfolded formulation of SFT, in which
all gauge symmetries (on the base manifold), including higher-spin
symmetries, are unbroken.

The affine realization of the singletons and its composites in the
case of $\widehat{\mso}(2,D-1)$ rests on the simple observation that,
for the \emph{subcritical level}\footnote{There is no coincidence
that $-k$ is the energy of the singleton ground state.
Interestingly, the subcritical case considered in
\cite{Dobrev:1990ng}, namely level $k'=-5/2$ in $D=4$, is dual to
our case in the sense that $k+k'=-h^{\vee}$. The precise relation
remains to spelled out.}
\bea k&=&-\epsilon_0\ \;\equiv\; -\frac{D-3}{2}\ ,\eea
there exists a singular vector at Virasoro level $2$ in the NS sector whose
decoupling plays the role of a non-perturbative equation of motion
\cite{Zamolodchikov:1986bd} selecting the physical operators. The
singular vector amounts to the affine \emph{hyperlight-likeness
condition}
\bea (M_A{}^C(z)M_{BC}(z))-\mbox{trace}&\sim&0\
,\label{hllintro}\eea
where by $M_{AB}(z)$ we denote the currents. In Section \ref{Sec:3}
we will discuss this condition in more detail and we will show that
the physical operators of the subcritical WZW model are
\emph{twisted primary fields}: the twist, labeled by an integer
$P\in\integ$, refers to that the lowest-weight conditions are
shifted upward and downward $P$ Virasoro levels for the KM charges
with positive and negative AdS energy, respectively. As we shall
see, the scalar singleton ($P=1)$ and anti-singleton $(P=-1$) and
their $|P|$-fold tensor products arise naturally in this
construction.

The singular vacuum vector induces a large number of singular
vectors in the various $P$-twisted Verma modules. The model
nonetheless contains non-unitary and non-topological states. Drawing
on the continuum limit of discretized branes \cite{Engquist:2005yt}
as well as on the sigma-model description of null surfaces in anti-de
Sitter spacetime \cite{Mikhailov:2004xw} (see also
\cite{Segal:1997ws,Gibbons:1999rb}), it makes sense to entertain
the idea that the coset model based on
\bea \frac{\widehat{\mathfrak{so}}(2,D-1)_{-\epsilon_0}}{
\widehat{\mathfrak{so}}(D-1)_{-\epsilon_0}\oplus
\widehat{\mathfrak{so}}(2)_{-\epsilon_0}}\ , \label{compactg} \eea
is a topological model consisting of symmetrized multipletons. In
\cite{Engquist:2005yt}, a specific continuum limit of a discretized
tensionless brane in $AdS_D$ was shown, using a phase-space
formulation on the $(D+1)$-dimensional ambient space, to be an
$Sp(2)$-gauged free-field model with critical dimension $D=7$. This
model was then argued to be dual to the
$\widehat{\mathfrak{so}}(6,2)_{-2}$ WZW model, and the gauging was
imposed by hand in order to remove unphysical states. We shall
describe the above gauging and its potential short-comings in more
detail in \cite{wip}.

In this paper we shall instead focus on the ungauged
$\widehat{\mathfrak{so}}(2,D-1)_{-\epsilon_0}$ model. Eventually, we
shall specify to the case of $\widehat\mso(2,3)_{-1/2}$ that can be
realized in terms of $4$ real symplectic bosons forming a quartet of
$\msp(4)$. The $2+2$ split of the bosons allows us to lift several
results from the study of the $\widehat\msp(2)_{-1/2}$ model in
\cite{Lesage:2002ch}, with the important exception of the
hyperlight-likeness condition \eq{hllintro}, which couples the two
doublets in a non-trivial fashion (the equation of motion of the
$\widehat\msp(2)_{-1/2}$ model is instead a $5$-plet at level $4$).

The paper is organized as follows: in Section \ref{Sec:2} we discuss
the hyperlight-likeness condition and the special role of the scalar
singletons in $D$ dimensions. In Section \ref{Sec:3} we lift this
condition to the non-compact WZW model and show how it is solved by
the twisted primary fields. In Section \ref{Sec:4} we present the
realization of the 4D model in terms of symplectic bosons and
multiple sets of Heisenberg oscillators. In Section \ref{Sec:5} we
compute the fusion rules using bosonization techniques. Finally, in
Section \ref{sec:6} we summarize our results.


\scs{Singletons and Multipletons}\label{Sec:2}                            


Singletons are ultra-short unitary representations of $\mso(2,D-1)$
that admit no flat spacetime limit. They were first discovered in
$D=4$ by Dirac in 1963 \cite{Dirac:1963ta}. His singletons may be
realized \cite{Dirac:1936fq,Mack:1969rr} as conformal particles on
the $4$-dimensional (singular) hypercone in $5$-dimensional
embedding space $\Real^{2,3}$ with signature $(--+++)$. Gauging the
world-line conformal group $Sp(2)$ leaves a $4$-dimensional physical
phase space parameterized by two Heisenberg oscillators. The $10$
oscillator bilinears generate a unitary representation of
$Sp(4)\simeq Spin(2,3)$ in the Fock space, which decomposes into a
scalar and a spinor representation consisting of the even and odd
states, respectively. Dirac seemed intrigued by the fact that his
representations do not survive the flat limit. This feature was
later stressed by Flato and Fr\o nsdal, who argued that, instead of
limiting the formulation of Quantum Field Theory to flat spacetime,
it might make more sense to start with a non-zero cosmological
constant and recover the flat case in a limit. Dirac's
representations could then play a role as hidden, or internal,
quantum variables. Indeed, a key property of AdS Quantum Field
Theory, discovered by Flato and Fr\o nsdal in \cite{Flato:1978qz},
is that massless and massive one-particle states are
\emph{composites} consisting of two or more singletons (they
introduced the term singleton referring to the fact that it occupies
a single line in the non-compact weight space). In other words, the
massless fermions and scalars, and even more interestingly, the
photons and gravitons, which one would consider as fundamental in
flat spacetime, are some form of two-singleton composites in AdS
spacetime. One recognizes a similar reasoning behind Thorn's string
bits and the more recent AdS/CFT correspondence. Flato and Fr\o
nsdal viewed, however, the singletons as ``gauge'' singlets as
opposed to the ``colored'' string bits and fundamental fields of the
holographic CFT's. One natural scenario in which singlet singletons
may arise is in the partonic description of extended objects in
$AdS_D$ advocated in \cite{Engquist:2005yt}. The main feature of
this proposal is to combine the discretization with a
$(D+1)$-dimensional phase-space formulation, without gauge fixing on
the worldvolume, and then argue for enhanced sigma-model gauge
symmetries in a combined tensionless and hypercone limit. The
resulting gauge group contains one $Sp(2)$ subgroup for each parton,
that hence can be identified as a singleton realized as a
$(D+1)$-dimensional conformal particle.


\scss{Elements of $\mso(2,D-1)$ Representation Theory}


To give a group-theoretical presentation of singletons and their
tensor products (see for example
\cite{Gunaydin:1981yq,Siegel:1988gd,Ferrara:2000nu,Angelopoulos:1999bz,Dolan:2005wy})
one starts from the hermitian $\mso(2,D-1)$ generators $M_{AB}$
($A,B=0',0,1,\dots,D-1$) obeying\footnote{We suppress Young
projections of indices, which are always of unit strength, unless
two sides of an equation have different index symmetries.}
\bea [M_{AB},M_{CD}]&=&4i\eta_{BC} M_{AD}\ ,\qquad M_{AB}\ =\
-M_{BA}\ =\ (M_{AB})^\dagger\ ,\eea
where $\eta_{AB}={\rm diag}(-,-,+,\cdots,+)$. In terms of the AdS
energy $E$ and translation-boost generators $L^\pm_r$
($r=1,\dots,D-1$), defined by
\bea E&=&M_{00'}\ =\ E^\dagger\ ,\qquad L_r^\pm\ =\ M_{r0'}\mp
iM_{r0}\ =\ (L^\mp_r)^\dagger\ ,\label{lpllm}\eea
and the spatial angular momenta $M_{rs}=-M_{sr}=(M_{rs})^\dagger$,
the commutation rules assume the following form
\bea [L^-_r,L^+_s]&=& 2\delta_{rs}E+2iM_{rs}\ ,\quad [E,L^\pm_r]\
=\ \pm L^\pm_r\ ,\quad [M_{rs},L^\pm_t]\ =\ 2i\delta_{st}L^\pm_r\
. \label{commrules}\eea
For $D\geq 4$ the physical representations are of lowest-weight
type\footnote{The cases $D=2$ and $D=3$ require a separate
analysis.}. To describe these representations one starts from the
Harish-Chandra module\footnote{Harish-Chandra modules are
generalized Verma modules where by definition all null states
generated by energy-lowering operators and spatial angular momenta
are factored out.} $\sfV(e_0,\mathbf m_0)$ (see
\emph{e.g.}~\cite{Ferrara:2000nu}) generated by the repeated action
of the energy-raising operators $L^+_r$ on a lowest-weight state
$\vert e_0,\mathbf m_0;\ell\rangle$ obeying
\bea L^-_r \vert e_0,{\mathbf m}_0;\ell\rangle\ =\ 0\ ,\qquad
(E-e_0)\vert{e_0,\mathbf m_0;\ell}\rangle\ =\ 0\ ,\eea
and with $\ell$ denoting the weights of the UIR of $\mso(D-1)$
labeled by the highest weight $\mathbf
m_0=(m_0^1,m_0^2,\dots,m_0^\nu)$, $m_0^1\geq m_0^2\geq \cdots\geq
|m_0^\nu|$, where $\nu=[(D-1)/2]$ and $m_0^\nu\geq 0$ if $D$ is
even integer. The spectrum of the energy operator $E$ acting in
$\sfV(e_0,\mathbf m_0)$ is bounded from below by the energy eigenvalue
$e_0$ of the lowest-weight state, which one sometimes refers to as
the ground state. The anti-linear inner product on the space of
ground states yields an anti-linear inner product on
$\sfV(e_0,\mathbf m_0)$ in its turn inducing a maximal invariant
subspace $\sfN(e_0,\mathbf m_0)\subset \sfV(e_0,\mathbf m_0)$.
This maximal ideal consists of null states, \emph{i.e.}~states
that are orthogonal to all states in $\sfV(e_0,\mathbf m_0)$. It
is nontrivial iff there exists at least one singular vector,
namely a state $\vert e'_0,\mathbf m'_0\rangle \in
\sfV(e_0,\mathbf m_0)$ with $e'_0>e_0$ obeying $L^-_r\vert
e'_0,\mathbf m'_0\rangle=0$. The singular vector generates its own
submodule of $\sfN(e_0,\mathbf m_0)$. There may be several null
submodules, possibly with additional substructure. Factoring out
$\sfN(e_0,\mathbf m_0)$ yields the non-degenerate lowest-weight
space
\bea \sfD(e_0,\mathbf m_0)&=& \sfV(e_0,\mathbf
m_0)/\sfN(e_0,\mathbf m_0)\ .\label{Dplus}\eea
Every lowest-weight space can be flipped ``upside down'' into a
highest-weight space,
\bea \sfD^-(e_0,\mathbf m_0)&=& \sfV^-(e_0,\mathbf
m_0)/\sfN^-(e_0,\mathbf m_0)\ ,\eea
built on the highest-weight state
\bea L^+_r \vert e_0,\mathbf m_0;\ell\rangle^-&=&0\ ,\qquad
(E+e_0)\vert{e_0,\mathbf m_0;\ell}\rangle^-\ =\ 0\ .\eea
We shall use the convention that superscript $+$ and $-$ indicate
lowest and highest-weight spaces and states, with $+$ set as
default value, and refer to the negative-energy states as
anti-states. Defining the linear map
\bea \pi\left(\vert e_0,\mathbf m_0\rangle^\pm\right)&=&\vert
e_0,\mathbf m_0\rangle^\mp\ ,\qquad \pi\left({}^\pm\langle
e_0,\mathbf m_0\vert\right)\ =\ {}^\mp\langle e_0,\mathbf m_0\vert\
,\eea
one can show that $\pi$ lifts to a linear involutive $\mso(2,D-1)$
automorphism given by
\bea \la{pimap} \pi(M_{ab})&=&M_{ab}\ ,\qquad \pi(P_a)\ =\ -P_a\
.\eea
where $M_{AB}=(M_{ab},P_a)$, with $P_a=M_{a0'}$, $a=(0,r)$ being the
space-time translations and $M_{ab}$ the Lorentz transformations. It
is also useful to introduce an anti-automorphism $\tau$ acting on
states and generators as follows
\bea \tau\left(\vert e_0,\mathbf
m_0\rangle^\pm\right)&=&{}^\mp\langle e_0,\mathbf m_0\vert\ ,\qquad
\tau(M_{AB})\ =\ -M_{AB}\ .\label{taumap}\eea
With these definitions, it follows that $\pi\tau=\tau\pi$.


\scss{Masslessness and Hyperlightlikeness}


From the commutation rules (\ref{commrules}) it follows that there
are no singular vectors if $e_0$ is large enough at fixed $\mathbf
m_0$, in which case $\sfD(e_0,\mathbf m_0)= \sfV(e_0,\mathbf m_0)$
is unitary and is referred to as a massive representation.
Lowering $e_0$ while keeping $\mathbf m_0$ fixed, singular vectors
appear for the first time at a critical value of $e_0$
\cite{Metsaev:1997nj}. In case $m_0^1\geq 1$, the singular vector
corresponds to a field-theoretic gauge artifact
\cite{Alkalaev:2006rw}, and one may refer to the critical
$\sfD(e_0,\mathbf m_0)$ as a massless representation. The simplest
case is $\mathbf m_0=(m_0^1)\equiv(m)$ with $m\geq 1$,
\emph{i.e.}~a ground state that is a symmetric rank $m$ tensor.
The corresponding unitary lowest-weight space is
\bea \sfD(m+2\epsilon_0,(m))\ ,\qquad \epsilon_0\ =\
\frac{D-3}{2}\ ,\eea
and the gauge modes are generated from the longitudinal
singular vector
\bea \vert m+2e_0+1,(m-1)\rangle_{r(m-1)}&=&\sum_{t=1}^{D-1}L^+_t
\vert m+2\e_0,(m)\rangle_{tr(m-1)}\  ,\eea
where we use the notation $r(m)\equiv r_1\cdots r_{m}$.

Interestingly, also scalar and spinor Harish-Chandra modules,
\emph{i.e.}~the cases $m=0,1/2$, exhibit critical behavior giving
rise to the \emph{scalar and spinor singletons}
\bea \sfD_0&\equiv& \sfD(\epsilon_0,(0))\ ,\qquad \sfD_{1/2}\
\equiv\ \sfD(\epsilon_0+1/2,(1/2))\ .\eea
The singular vectors are now given by
\bea L^+_rL^+_r\vert \epsilon_0,(0)\rangle\ ,\qquad
(\gamma_r)_\a{}^\b L^+_r\vert \epsilon_0+1/2,(1/2)\rangle_\beta\
,\eea
where $(\gamma_r)_\a{}^\beta$ are Dirac matrices (squaring the Dirac
operator one finds that $L^+_rL^+_r\simeq 0$ also for the spinor).
The singletons therefore consist of single discrete lines
$\{n+\epsilon_0,(n)+\mathbf m_0\}_{n=0}^\infty$ in the non-compact
weight space. In the Harish-Chandra module, the singular vectors are
compact, or Bogoliubov-transformed, versions of the
$(D-1)$-dimensional equations of motion of conformal scalar and
spinor fields \cite{Gunaydin:1998jc}. There are also unitary
singleton-like representations with $m\geq 1$, corresponding to
discrete subcritical values of $e_0$, but we shall not be interested
in them here.

Alternatively, the quantum-mechanical equation of motion of the
singleton can be written directly in terms of the angular momenta
$M_{AB}$ on a manifestly $(D+1)$-dimensionally covariant form as
the following \emph{hyperlight-likeness
condition}\footnote{Realizing the angular momenta as
$M_{AB}=X_AP_B-X_BP_A$ where $X^A$ and $P^B$ are the coordinates
and momenta of the conformal particle in $\Real^{2,D-1}$, one can
show that $V_{AB}$ is proportional to the $Sp(2)$ generators
$X^2$, $P^2$ and $\{X^A,P_A\}$, so that $V_{AB}\simeq 0$ comprises
the light-likeness condition $P^2\simeq 0$ as well as the
hyper-cone condition condition $X^2\simeq 0$. The
hyperlight-likeness condition forces the $D$-dimensional
space-time energy-momentum into rotational motion rather than
ordinary geodesic, \emph{e.g.}~light-like, motion.},
\bea \langle\Psi \vert V_{AB}\vert \Psi'\rangle&=&0\
,\label{hll}\eea
where $V_{AB}$ is the traceless operator
\bea V_{AB}&=& \frac12 M_{(A}{}^C M_{B)C}-{\eta_{AB}\over D+1}C_2\
,\qquad C_2\ =\ \frac12 M^{AB}M_{AB}\ .\label{VAB1}\eea
In a lowest or highest-weight space, this is equivalent to that
the ground state $\vert\Omega\rangle$ obeys
\bea V_{rs}\vert{\Omega}^\pm\rangle&=& 0\ .\label{Vrs}\eea
In the case of scalar and spinor lowest-weight spaces in $D\geq
4$, one can show that the hyperlight-likeness condition holds only
for scalar singletons in $D\geq 4$ and the spinor singleton in
$D=4$, \emph{i.e.}\footnote{One can also show that the scalar and
spinor singletons in $D=3$ are hyperlight-like.}
\bea \la{hlx1} \sfD^\pm_0&:&\quad \mbox{hyperlight-like for all $D$}\ ,\\
\la{hlx2} \sfD^\pm_{1/2}&:&\quad \mbox{hyperlight-like iff
$D=3,4$}\ . \eea
We note that in the singleton Harish-Chandra module $V_ {AB}\vert
\chi\rangle$ are null states for arbitrary $\vert\chi\rangle$. In
particular, the $(D-1)$-dimensional mass-shell condition is
recovered from $V^{++}=\frac12 L^+_r L^+_r$, where $X^\pm=X_{0'}\mp
iX_0$. Moreover, acting on $\vert \e_0+n,(m)\rangle\in
\sfV(\e_0,(0))$ with $m=n-2k$, $k=1,2,...$ (\emph{i.e.}~containing
$k$ factors of $L^+_r L^+_r$) with $V^+_r$ and decomposing into
symmetric traceless and trace parts, yields
\bea V^+_{\{r}\vert \e_0+n,(m)\rangle_{r(m)\}}&=& ikL^+_{\{r}
\vert\e_0+n,(m)\rangle_{r(m)\}}\
,\\V^+_s\vert\e_0+n,(m)\rangle_{sr(m-1)}&=&\frac{i}2L^+_r\vert\e_0+n,(m)\rangle_{rs(m-1)}\
,\eea
where $\{r(m)\}$ denotes the traceless part of $r(m)$. Each of
these equations separately implies that all states in the
Harish-Chandra module with $k>0$ are null states.


\scss{Compositeness and Higher-Spin Algebra}


The fundamental nature of the singletons was first exhibited in
$D=4$ by Flato and Fr\o nsdal in \cite{Flato:1978qz}, where they
derived the following decomposition under $\mso(2,3)$ of the tensor
product of two scalar singletons,
\bea \sfD_0\otimes \sfD_0&=&\bigoplus_{m=0,1,2,\dots}\sfD(m+1,m)\
,\label{FF1}\\[4pt] \sfD_0\otimes
\sfD_{1/2}&=&\bigoplus_{2m=1,3,\dots} \sfD(m+1,m)\
,\label{FF2}\\[4pt] \sfD_{1/2}\otimes \sfD_{1/2}&=& \sfD(2,0)\oplus
\bigoplus_{m=1,2,\dots}\sfD(m+1,m) \label{FFtheorem}\ ,\eea
where we note that all representations on the right-hand sides are
massless. Their result generalizes straightforwardly to
two-singleton composites $D$ dimensions
\cite{Angelopoulos:1997ij,Sezgin:2001zs,Sezgin:2001yf,Sezgin:2001ij,Vasiliev:2004cm,Engquist:2005yt}
\bea \sfD_0\otimes
\sfD_0&=&\bigoplus_{m=0,1,2,\dots}\sfD(m+2\epsilon_0,(m))\
.\label{FFD}\eea
For this reason, one refers to the symmetric traceless rank $m\geq
0$ tensors in $D$ dimensions and the pseudo-scalar $\sfD(2,0)$ in
$D=4$ as composite massless\footnote{In $D=4$ the composite massless
representations admit extensions to the conformal group whose
restrictions to the Poincar\'e group are ordinary massless particles
with helicity $\pm m$. These can be compared with the conformally
coupled scalar in $D$ dimensions, that has $e_0=(D-1\pm1)/2$, which
can equal the composite value $e_0=D-3$ only in $D=4$ and $D=6$, and
that of the composite pseudo-scalar value $e_0=D-2$ only in $D=2$
and $D=4$.}. The two-singleton composites can be decomposed under
$\cS_2$ into symmetric and anti-symmetric Young
projections\footnote{It is intriguing that
$[\sfD_0\otimes\sfD_0]_{\rm S}$ agrees with
$[\sfD_{1/2}\otimes\sfD_{1/2}]_{\rm AS}$ except in the scalar
sector, where the former contains the parity-invariant scalar
$\sfD(1,0)$ and the latter the pseudo-scalar $\sfD(2,0)$. This
points via holography to some form of bose-fermi correspondence in
three-dimensional conformal field theory
\cite{Leigh:2003gk,Sezgin:2003pt}.}
\bea [\sfD_0\otimes \sfD_0]_{\rm S} &=&
\bigoplus_{m=0,2,4,\dots}\sfD(m+2\epsilon_0,(m))\ ,\\[4pt]
{}[\sfD_0\otimes \sfD_0]_{\rm AS}
&=&\bigoplus_{m=1,3,5,\dots}\sfD(m+2\epsilon_0,(m))\ .\eea
Adhering to the nomenclature introduced in \cite{Beisert:2004di}, we
shall refer to Young-projected multi-singleton composites as
\emph{multipletons}, and in particular to the massless cases as
symmetric and anti-symmetric doubletons\footnote{The doubletons of
$\mso(2,3)$, which are massless conformal tensors in four
dimensions, are singletons of $\mso(2,4)\simeq \msu(2,2)$. This
particular notion of doubletons was first introduced in
\cite{Gunaydin:1984fk}.}. In general, the tensor product of $P\geq
2$ singletons decomposes under $\cS_P$ into $P!$ $P$-tupletons. For
$P>2$ these decompose under $\mso(2,D-1)$ into massive
representations, although as far as we know the explicit form of
this decomposition has not been given in the literature.

The scalar singleton $\sfD^\pm_0$ is the fundamental representation
of the minimal bosonic higher-spin algebra $\mhso_0(2,D-1)$. This
algebra is an infinite-dimensional Lie-algebra extension of
$\mso(2,D-1)$, defined by factoring out the annihilator of
$\sfD^\pm_0$ from the $Env(\mso(2,D-1))$
\cite{Vasiliev:2004cm,Sagnotti:2005ns,Engquist:2005yt}\footnote{The
minimal higher-spin algebra is denoted by $ho(1|2;[2,D-1])$ in
\cite{Vasiliev:2004cm}.}. The algebra $\mhso_0(2,D-1)$ acts
transitively on $\sfD^\pm_0$ (and $\sfD^\pm_{1/2}$ in $D=4$) and
irreducibly on the multipletons. The irreducibility of the
multipletons, \emph{i.e.} the absence of non-trivial
$\mhso_0(2,D-1)$-invariant tensors in $(\sfD^\pm_0)^{\otimes P}$, is
a consequence of the fact that $\mhso_0(2,D-1)$ contains $\msu(k)$
subalgebras for unbounded values of $k$. These subalgebras are
obtained by consistently truncating to generators that have
non-vanishing matrix elements only between singleton states with
excitation energies $\e_0+n$ for $n=0,1,\dots,N$ ($k$ can be made
arbitrarily large by taking $N$ sufficiently large). As an aside,
for later reference, we note that both $(\mhso_0(2,D-1),\sfD^+_0)$
and $(\mhso_0(2,D-1),\sfD^-_0)$ are formal $N\rightarrow \infty$
limits of $(\msu(N),N)$. Thus, apart from the standard
$\mhso_0(2,D-1)$-invariant inner product
$\langle\cdot\vert\cdot\rangle:\sfD^\pm_0\otimes
\sfD^\pm_0\rightarrow\Comp$, there also exists an
$\mhso_0(2,D-1)$-invariant \emph{bilinear} inner product
$\left(\cdot,\cdot\right):\sfD^\pm_0\otimes
\sfD^\mp_0\rightarrow\Comp$ defined by
\bea
\left(\lambda\vert\Psi\rangle,\lambda'\vert\Psi'\rangle\right)&=&
\lambda\lambda'\left(\vert\Psi\rangle,\vert\Psi'\rangle\right)\
,\\\left(\vert\Psi\rangle,M_{AB}\vert\Psi'\rangle\right)&=&
\left(-M_{AB}\vert\Psi\rangle,\vert\Psi'\rangle\right)\
,\\\left(\vert\epsilon_0,(0)\rangle^-,\vert\epsilon_0,(0)\rangle^+\right)&
=&\left(\vert\epsilon_0,(0)\rangle^+,\vert\epsilon_0,(0)\rangle^-\right)\
=\ 1\ .\label{bilin}\eea

The higher-spin algebra, by its construction, contains an
infinite-dimensional Cartan subalgebra, so that its representation
theory falls outside that of Lie algebras with finite-dimensional
Cartan matrices. On the other hand, from a physical point-of-view,
the higher-spin algebra appears to be too small to unify interactions that
mix multipletons with different values of $P$ (with the exception of
(classically consistent truncations to) self-interactions in the
massless $P=2$ sector). As found in \cite{Engquist:2005yt},
multipletons arise in a discretized description of tensionless
extended objects in anti-de Sitter spacetime. It was further argued
that the continuum limit leads to a non-compact WZW model, wherein
the singletons arise as twist-fields in the R sector\footnote{In
\cite{Engquist:2005yt} we used the term spin-field, but here we
shall instead use the more commonly used term twist-field.}, the
tensoring is lifted to fusion, and the role of the higher-spin
algebra is replaced by an affine extension of $\mso(2,D-1)$.


\scs{On Subcritical $\widehat{\mso}(2,D-1)$ WZW Models}\label{Sec:3}


In this section we propose an $\widehat{\mso}(2,D-1)$ WZW model at the
subcritical level $-(D-3)/2$, which we claim accommodates the scalar
singleton as well as all its composites. Here, our approach is
purely algebraic, based on using spectral flow to solve the
current-algebra version of the hyperlight-likeness condition
\eq{hll}. The construction will then be explored in more detail in
the case of $D=4$ in the coming sections using symplectic bosons and
free-field realizations. The affine representation spaces contain
many unphysical states over and above the desired singletons and
multipletons, which one may think of as generalized ground states.
Whether a unitary model can be extracted by some form of gauging is
still an open issue, and we refer to \cite{wip} for more details.


\scss{Affine Hyperlightlikeness}


In what follows we shall focus on the holomorphic sector of the
WZW model. The underlying $\widehat{\mso}(2,D-1)_k$ currents obey
the operator product expansion
\bea M_{AB}(z) M_{CD}(w)&=&{2k\eta_{AC}\eta_{BC}\over
(z-w)^2}+{4i\eta_{BC}M_{AD}(w)\over z-w} + {\rm finite}\ ,
\label{MOPE}\eea
where the level $k$ is chosen such that the normal-ordered
field\footnote{The normal ordering operation is defined by
$(AB)(z)=\oint_z {dx\over 2\pi i}{A(x)B(z)\over x-z}$ (for example,
see \cite{DiFrancesco:1997nk}).}
\bea V_{AB}(z)&=&\frac12 \left( M_{(A}{}^C M_{B)C}\right)(z)-{\rm
trace}\ ,\label{VABz}\eea
is a singular vector in the NS sector. In other words, $k$ is
fixed such that $V_{AB}(z)$ is a Kac-Moody primary obeying
\bea M_{AB}(z)V_{CD}(w)&=& {4i\eta_{BC}V_{AD}(z)\over z-w}+ {\rm
finite}\ .\label{MV}\eea
Since $V_{AB}$ is an $\mso(2,D-1)$ tensor with highest weight $(2)$
and canonical conformal weight $2$, the Sugawara construction
implies that $V_{AB}$ can be a KM primary iff
\bea h[V_{AB}]&\equiv&{C_2[\mso(2,D-1)|(2)]\over 2(k+h^\vee)}\ =\
{2(2+D-1)\over 2(k+D-1)}\ =\ 2\ ,\eea
from which we read off the \emph{subcritical level}
\bea k&=&-\epsilon_0\ .\eea
Indeed, for this value the double contractions in
$M_{AB}(z)V_{CD}(w)$ cancel, and one is left with \eq{MV}. The
Sugawara stress tensor becomes
\bea T(z)&=& {1\over 2(D+1)}(M^{AB}M_{AB})(z)\label{sugawara}\eea
with central charge $c=-D(D-3)/2$.
The hermitian conjugation is given by $X^\dagger=((X)^\ast)^\tau$
where $*$ is an anti-linear automorphism $*$ and $\tau$ a linear
anti-automorphism (BPZ conjugation), defined by
\bea \left(M_{AB}(z)\right)^*&=&-M_{AB}(z)\ ,\qquad
(M_{AB}(z))^\tau\ =\ -\frac{1}{z^2}M_{AB}\left(z^{-1}\right)\ ,\eea
resulting in
\bea \la{conjgen} (M_{AB,n})^*&=&-M_{AB,n}\ ,\qquad (M_{AB,n})^\tau\
= \ -M_{AB,-n}\ ,\qquad (M_{AB,n})^\dagger\ =\ M_{AB,-n}\
.\qquad\eea
Acting on the 0-modes, the BPZ conjugation reduces to the
anti-automorphism $\tau$ defined in \eq{taumap}. The lift of the
$\mso(2,D-1)$ automorphism $\pi$ given in \eq{pimap} to the affine
case is given by
\bea \la{pimapaffine1} \pi(M_{ab,n})&=&M_{ab,n}\ , \qquad
\pi(P_{a,n})\ = \ -P_{a,n} \ .\eea

Turning to the primary fields of the WZW model, the decoupling of
the singular vacuum vector at the level of three-point functions
forces the primaries to obey an \emph{equation of motion}
\cite{Zamolodchikov:1986bd,Lesage:2002ch}, which is a necessary
condition that fixes the admissible representations but not their
multiplicities (a complete determination of the spectrum requires
further input from demanding closure of the operator product
expansion, crossing symmetry and modular invariance). Denoting the
primary fields and states by $V_\Lambda(z)$ and
$\vert\Lambda\rangle=V_\Lambda(0)\vert 0\rangle$, and the
corresponding KM Verma modules by $\widehat\sfV(\Lambda)$, the
decoupling amounts to
\bea \langle\Psi\vert V_{AB,n} \vert \Psi'\rangle&=& 0\quad\mbox{for
all $n\in \integ$ and $\vert \Psi\rangle,\ \vert \Psi'\rangle\in
\widehat \sfV(\Lambda)$}\ ,\label{decoupl}\eea
where the inner products are defined using the hermitian conjugation
\eq{conjgen}. The decoupling \eq{decoupl} is equivalent to
$\langle\Lambda\vert V_{AB,n} \vert \Lambda\rangle=0$ for all $n$.
Assuming that $\langle\Lambda\vert$ has a fixed $L_0$ eigenvalue,
this is the same as $\langle\Lambda\vert V_{AB,0} \vert
\Lambda\rangle=0$. Finally, assuming that $\langle\Lambda\vert$ is a
$\mso(2,D-1)$ ground state, one finds
\bea V_{rs,0}\vert \Lambda\rangle&=&0\ ,\label{zzeq}\eea
which is the affine version of \eq{Vrs}.

In general, the primary field may be \emph{twisted} in the sense
that $M_{AB,n}\vert\Lambda\rangle$ vanishes for some $AB$ and $n<0$
and does not vanish for some $AB$ and $n>0$. In our case, the
relevant twisting is described using the compact basis, \emph{viz.}
\bea [L^-_{r,m},L^+_{s,n}]&=&
2iM_{rs,m+n}+2\delta_{rs}E_{m+n}+2\epsilon_0
m\delta_{rs}\delta_{m+n,0}\ ,\label{KMCR1}\\[4pt]
{}[E_m,E_n]&=&-\epsilon_0m\delta_{m+n,0}\ ,\label{KMCR2}\\[4pt]
{}[M_{rs,m},M_{tu,n}]&=&
4i\delta_{st}M_{ru,m+n}-2\epsilon_0m\delta_{rt}\delta_{su}\delta_{m+n,0}\
.\label{KMCR3}\eea
Given an integer $P$, a \emph{$P$-twisted primary
field}\footnote{Primary fields of this type are also referred to
in the literature (see
\emph{e.g.}~\cite{Maldacena:2000hw,D'Appollonio:2003dr}) as
spectrally flowed primary fields.} $V_{[P];\Lambda}(z)$ is by
definition taken to obey
\bea L^\pm_{r,n}\vert [P];\Lambda\rangle&=&0\quad\mbox{for
$n\geq\pm P+1$}\ ,\label{TW1}\\[4pt] M_{rs,n}\vert
[P];\Lambda\rangle&=&E_n\vert[P];\Lambda\rangle\ =\
0\quad\mbox{for $n\geq 1$}\ ,\label{TW2}\eea
and to be a ground state with lowest weight labels
$(e^{[P]}_0,{\mathbf m}^{[P]}_0)$ of a Harish-Chandra module of the
$\mso(2,D-1)$ algebra with generators
\bea L^{[P]\pm}_{r}&=& L^\pm_{r,\pm P}\ ,\qquad M^{[P]}_{rs}\ =\
M_{rs,0}\ ,\qquad E^{[P]}\ =\ E_0-P\e_0\ .\label{TW3}\eea
We note that from $(E^{[P]}-e^{[P]}_0) |[P],\Lambda\rangle = 0$ it
follows that
\bea e_0&=& e^{[P]}_0 + P\e_0\ ,\qquad {\mathbf m}_0\ =\ {\mathbf
m}_0^{[P]}\ .\eea
We shall label the $P$-twisted ground states by
\bea \Lambda&=& (h;e_0,{\mathbf m}_0)\ ,\eea
where $h$ is the conformal weight. The action of the KM creation
operators on $\vert[P];h;e_0,{\mathbf m}_0\rangle$ generates Verma
modules $\widehat\sfV_{[P];h}(e_0,{\mathbf m}_0)$ containing null
submodules $\widehat\sfN_{[P];h}(e_0,{\mathbf m}_0)$ generated by
$P$-twisted singular vectors (containing at least all excitations
generated by the modes of $V_{AB}$). Factoring out the null states
yields the twisted lowest weight spaces
\bea \widehat\sfD_{[P];h}(e_0;\mathbf m_0) &=&
{\widehat\sfV_{[P];h}(e_0;\mathbf m_0)\over
\widehat\sfN_{[P];h}(e_0;\mathbf m_0)}\ .\eea
We note that the standard definition of a KM weight space is
recovered for $P=0$. For $P=\pm 1$, one additional singular vector
has been factored out from the Verma module, namely
$L^{\mp}_{r,-1}\vert[\pm1 ];\Lambda\rangle$.


\scss{Twisted Primary Scalars and Multipletons}


Let us consider the special case of a $P$-twisted primary field
$V_{[P]}(z)$ that is a \emph{singlet} under $\mso(2,D-1)_{[P]}$,
\emph{i.e.}
\bea L^{\pm}_{r,n}\vert [P]\rangle &=& 0\ ,\qquad n\geq \pm P\
,\label{twpr1}\\[4pt] (E_n-\delta_{n0}P\epsilon_0)\vert [P]\rangle&=&0\
,\qquad M_{rs,n}\vert [P]\rangle\ =\ 0\ ,\qquad n\geq 0\
.\label{twpr2}\eea
These states are also conformal primaries,
\bea (L_n-h_{[P]}\d_{n0})\vert[P]\rangle&=&0\ ,\qquad h_{[P]}\ =\
-{P^2\epsilon_0\over 2}\  ,\label{hP}\eea
where $n\geq 0$ and $L_n$ are the Virasoro generators taken from the
Sugawara stress tensor \eq{sugawara}. Since $\vert [P]\rangle$ is an
$SO(D-1)$ scalar, it follows that
\bea V_{rs,0}\vert[P]\rangle&=&{1\over
D-1}\delta_{rs}V_{tt,0}\vert[P]\rangle\ .\eea
By explicit calculations one can then go on to show that
$V_{rr,0}\vert[P]\rangle=0$, \emph{i.e.}~the twisted-primary
singlets defined above obey the equations of motion \eq{zzeq},
\emph{i.e.}
\bea V_{rs,0}\vert [P]\rangle&=&0\ .\eea
The scalar twisted lowest-weight spaces, that we shall denote by
\bea \widehat\sfD_{[P]}&=& \widehat \sfD_{[P],-{P^2\e_0\over
2}}(P\e_0,(0))\ ,\eea
contain negative norm states, which should be removed carefully by
imposing suitable gauge conditions \cite{Dobrev:1990ng,wip}. In this
paper we shall not enter this important discussion. Instead, let us
highlight the interesting, potentially physical, subspace
$\sfD_{[P]}\subset \widehat\sfD_{[P]}$ given by
\bea \sfD_{[P]}&=&\bigoplus_{e_0,\mathbf
m_0}\sfD_{[P];h}(e_0,\mathbf m_0)\ ,\label{DP}\eea
where $\sfD_{[P];h}(e_0,\mathbf m_0)$ are defined to be the
(untwisted) $\mso(2,D-1)$ weight spaces generated by acting with
$L^\sigma_{r,0}$, with $\sigma={\rm sign}(P)$, on ground states of
the form
\bea \vert [P];h;e_0,\mathbf m_0\rangle^{\sigma}
&=&\left(\prod_{r,|n|\leq |P|-1 }
L^\sigma_{r,n}\right)\vert[P]\rangle + \vert[P];\Psi\rangle\ ,\eea
where the state $\vert[P];\Psi\rangle$ contains at least one
excitation by one of the compact subalgebra generators
$\left\{E_{-n},M_{rs,-n}\right\}_{n=0}^{|P|-1}$, and is determined
by
\bea L^{-\sigma}_{r,0}\vert [P];h;e_0,\mathbf
m_0\rangle^{\sigma}&=&0\ .\eea
We note that the weight spaces in $\sfD_{[P]}$ are of lowest-weight
type for $P>0$ and highest-weight type for $P<0$, and we shall
therefore drop the superscript $\sigma$ on the states. In
particular,
\bea \vert[P]\rangle&=& \vert[P];-{P^2\epsilon_0\over
2};2\sigma\e_0,(0)\rangle\ \la{agsofws} \eea
are ground states of scalar weight spaces. The state
$\vert[0]\rangle$ is the NS vacuum, so that
\bea \sfD_{[0]}&=& \sfD_{[0]}(0,0)\ .\eea
For $P=\pm 1$, we identify $\vert[\pm1]\rangle$ as the ground states
of the scalar singleton, $\sfD^+=\sfD_{[1];-\e_0/2}(\e_0,(0))$, and
anti-singleton, $\sfD^-=\sfD_{[-1];-\e_0/2}(-\e_0,(0))$, and hence
\bea \sfD_{[\pm 1]}&\simeq&  \sfD^\pm_0\ .\eea
For $P=\pm 2$, we identify $\vert[2]\rangle$ and $\vert[-2]\rangle$
as the ground states of the composite massless scalar and
anti-scalar, respectively. Acting on $\vert [\pm 2]\rangle$ with traceless
strings of $L^+_{r,1}$ generators yields the \emph{composite massless
higher-spin ground states} ($m=0,1,2,3,\dots$)
\bea \vert [\pm2];-2\e_0-m;\pm(m+2\e_0),(m)\rangle_{\{r(m)\}}\ =\
L^\pm_{\{r_1,1}\cdots L^\pm_{r_{m}\},1}\vert[\pm2]\rangle\
.\label{massless}\eea
We note that the trace parts sit in $\widehat\sfN_{[\pm
2];-2\e_0-m}(\pm(m+2\e_0),(m))$. Similarly, traceless strings of
$L^\sigma_{r,-1}$ generators dressed up with suitable correction
terms involving subalgebra generators provide massless ground states
of the form $\vert[\pm 2];-2\e_0+m; m+2\e_0,(m)\rangle$. For
example, the photon ground state with conformal weight $-2\e_0+1$ is
given by
\bea \vert[2];-2\e_0+1; 1+2\e_0,(1)\rangle_r&=&
\left(L^+_{-1,r}+{i\over
1+2\e_0}M_{rs,-1}L^+_{s,0}\right)\vert[2]\rangle\
.\label{photon}\eea
Similar constructions can be given for higher spin and for $P=-2$.
Thus
\bea \sfD_{[\pm 2]}&\supset & \sfD_{[\pm2];-2\e_0}(\pm
2\e_0,(0))~\oplus~\sfD'_{[\pm 2];+}~\oplus~ \sfD'_{[\pm 2];-}\
,\label{sfDpm2}\eea
where we have defined
\bea\la{ml1} \sfD'_{[\pm 2];-}&=&\bigoplus_{m=1,2,...} \sfD_{[\pm
2];-2\e_0-m}(\pm(m+2\e_0),(m))\ ,\\[4pt] \la{ml2}\sfD'_{[\pm 2];+}&=&
\bigoplus_{m=1,2,...} \sfD_{[\pm 2];-2\e_0+m}(\pm(m+2\e_0),(m))\
.\eea
From \eq{FFD} it follows that
\bea \sfD_{[\pm2];-2\e_0}(\pm
2\e_0,(0))\oplus\sfD'_{[\pm2];-}&\simeq& \sfD^\pm_0\otimes
\sfD^\pm_0\ .\eea
We expect the above pattern to generalize, so that
\bea \sfD_{[P]}&\supset& \sfD_{[P];-}\ \simeq\
(\sfD^\sigma_0)^{\otimes |P|}\ ,\label{DPminus}\eea
where $\sfD_{[P];-}$ is defined to be the space of states in
$\sfD_{[P]}$ with minimal $L_0$-eigenvalue for fixed $\mso(2,D-1)$
quantum numbers.

The above analysis suggests that the scalar twisted primary states
$|[P]\rangle$ correspond to scalar twisted primary fields $V_{[P]}$
obeying the simple fusion rule $V_{[P]}\times V_{[P']}=V_{[P+P']}$.


\scss{Spectrum-Generating Flow and Fusion}


Spectral flow is an operation which shifts the modes of affine
generators such that the spectrally-flowed algebra is isomorphic
to the original one. It is known that spectral flow in WZW models
based on affine Lie algebras with infinite-dimensional zero-mode
representations connects an infinite set of sectors which are all
required for the consistency of the model, see {\emph
e.g.}~\cite{Gaberdiel:2001ny,Maldacena:2000hw,Lesage:2002ch,D'Appollonio:2003dr}.
In our case, these sectors are labeled by the integer $P$,
containing as ``zero modes'' states in the $|P|$th tensor product
of singletons ($P>0$) or antisingletons ($P<0$).

The possible spectra of scalar twisted primaries is restricted by a
delicate interplay between fusion and spectral flow. At the level of
the fusion rules, which we denote by $\times$ and whose entries are
affine representation labels $\Lambda$, the spectral flow operation,
\emph{viz.}
\bea \O_P[V_\Lambda]&=& V_{\O_P(\Lambda)}\ ,\qquad P\in\integ\
,\eea
can be composed as follows \cite{Gaberdiel:2001ny}
\bea \O_P\circ \O_{P'}&=&\O_{P+P'}\ .\label{Gab1}
\eea
By using the commutativity of fusion rules, one can also show that
\bea
\O_{P}[V_\Lambda\times V_{\Lambda'}]&=&\O_P[V_\Lambda]\times
V_{\Lambda'}\ =\ V_\Lambda\times \O_{P}[V_{\Lambda'}]\
.\label{Gab}\eea
We note that as a consequence $\O_{P+P'}[V_\Lambda\times
V_{\Lambda'}]=\O_P[V_\Lambda]\times \O_{P'}[V_{\Lambda'}]$. At the
level of the operator product algebra the distribution of spectral
flows induces automorphisms,
\bea \O_P[A(z)B(w)]&=& (\O_P A\O_P^{-1})(z) (\O_P[B])(w)\ .\eea
The spectrally flowed currents are defined by
\bea (\O_P L^\pm_r\O_P^{-1})(z) &=& z^{\pm P}L_r^\pm(z)\ ,\qquad
(\O_P M_{rs} \O_P^{-1})(z) \ =\   M_{rs}(z)\ ,
\\[4pt](\O_P E \O_P^{-1})(z) &=& E(z)-P\epsilon_0 z^{-1}\ ,
\label{flowE}\eea
or in terms of the charges
\bea \O_P L^\pm_{r,n} \O_P^{-1}&=& L^\pm_{r,n\pm P}\ ,\qquad \O_P
M_{rs,n} \O_P^{-1} \ =\ M_{rs,n}\ ,\\[4pt] \O_P E_n \O_P^{-1} &=&
E_n-P\epsilon_0\delta_{n,0}\ .\label{flowEn}\eea
The shift in the energy is compatible with the central extension in
\eq{KMCR1}. The spectral flow of the Sugawara stress tensor
\eq{sugawara} is by definition given by
\bea (\Omega_P T \Omega_P^{-1})(z)&=& \oint_z {dx\over 2\pi
i(x-z)} (\Omega_P M^{AB} \Omega_P^{-1})(x) (\Omega_P
M_{AB}\Omega_P^{-1})(z)\ .\label{OPsugawara}\eea
Expanding the operator product and evaluating the residue at $x=z$
one finds
\bea (\O_P T \O_P^{-1})(z)&=& T(z)+PE(z)-{P^2\e_0\over 2}z^{-2}\
,\label{flowT}\eea
or, at the level of Virasoro generators,
\bea \O_P L_n\O_P^{-1}&=&L_n+PE_n-{P^2\e_0\over 2}\delta_{n,0}\
,\label{flowTn}\eea
where $\O_P L_n\O_P^{-1}$ indeed obeys the same Virasoro algebra as
$L_n$. The mixing between $T(z)$ and the AdS-energy current $E(z)$
can also be expressed as
\bea E(z)&=&(\O_P E \O_P^{-1})(z)+P\e_0 z^{-1}\ ,\label{flowE1}\\[4pt]
T(z)&=&(\O_P T
\O_P^{-1})(z)-P(\O_P E \O_P^{-1})(z)z^{-1}-{P^2\e_0\over 2}z^{-2}\
.\label{flowT1}\eea

A family of $\mso(2,D-1)_{[P]}$-singlet $P$-twisted primary fields,
obeying \eq{twpr1} and \eq{twpr2}, can now be constructed by
applying spectral flow to the identity,
\bea V_{[P]}&=& \O_P[\id]\ ,\qquad P\in \integ\ .\eea
By construction these fields obey the hyperlight-likeness condition
\eq{zzeq}, which one can also verify directly by calculating
\bea \O_P V_{rs,0} \O_P^{-1}&=&V_{rs,0}\ ,\eea
which implies $V_{rs,0}\vert[P]\rangle=\O_P[V_{rs,0}\vert
0\rangle]=\O_P[0]=0$. We note that $\O_P E_n\O_P^{-1}
|[P]\rangle=\O_P L_n\O_P^{-1} |[P]\rangle=0$ for $n\geq 0$ is
consistent with \eq{flowEn} and \eq{flowTn}, and that the operators
on the left and right-hand sides of \eq{flowE1} and \eq{flowT1} are
normal-ordered with respect to the $0$-twisted and $P$-twisted
ground states, respectively. The various conjugates of the twisted
primary states are given by
\bea \la{conj1} |[P]\rangle^*\ = \ |[-P]\rangle\ , \qquad
|[P]\rangle^\tau\ = \ \langle[P]|\ ,\qquad \pi(|[P]\rangle\ =\
|[-P]\rangle\ ,\eea
implying the hermitian conjugates
\bea \la{conj2} |[P]\rangle^\dagger\ \equiv \ (|[P]\rangle^*)^\tau\
= \ \langle[-P]|\ . \eea
We can thus define an inner product by declaring
\bea \la{normsP} \langle [P']|[P]\rangle&=&\d_{P+P',0}\ , \qquad
P,P'\in\integ\ . \eea
By definition $\langle[P]\vert
M_{AB,n}=((M_{AB,n})^\dagger|[-P]\rangle)^\dagger$, which implies
that the above definition is consistent with the assignment of $E_0$
eigenvalues. We note that there also exists a bilinear inner product
$(\cdot,\cdot)$, induced by the two-point functions, \emph{viz.}
\bea
\langle0|O_{\Psi}(z)O_{\Psi'}(w)|0\rangle&=&{(|\Psi\rangle,|\Psi'\rangle)\over
(z-w)^{2h(\Psi)}}\ ,\eea
where $O_\Psi(z)$ denotes the vertex operator corresponding to the
state $|\Psi\rangle$. This bilinear inner product reduces to
\eq{bilin} upon restricting to the set of $[\pm]$-states generated
by the 0-modes.

The composition rule \eq{Gab1} and distribution rule \eq{Gab}
imply the simple fusion rule\footnote{In the case of WZW models
based on affine algebras with infinite-dimensional zero-mode
representations, one has to pay extra attention to the fusion
rules. For instance, the Verlinde formula does not necessarily
apply, {\it cf.}~\cite{Lesage:2002ch} and references therein.}
\bea \la{fusprod} V_{[P]}\times V_{[P']}&=& V_{[P+P']}\ .\eea
The analysis so far suggests an $\widehat{\mso}(2,D-1)_{-\e_0}$ model
with spectrum given by
\bea \widehat \sfD&=& \bigoplus_{P\in\integ} \widehat
\sfD_{[P];-{P^2\e_0\over 4}}(P\e_0,(0))\ .\label{WZWspectrum}\eea
Clearly, the closure of the fusion rules is only a necessary
criterion, and we leave a more detailed study of this model,
\emph{e.g.} the properties under modular transformations and
locality of the operator product expansion, to future work. In fact,
already in the case of $D=4$, we observe that the above assertion
does not take into account the hyperlight-likeness of the spinor
singleton, \emph{cf.}~\eq{hlx2}. As we shall show next, the affine
extension of this representation can be added and spectrally flowed
leading to an \emph{extended} 4D model with additional
tensor-spinors.


\scss{Twisted Primary Spinors in $D=4$}


The case of $D=4$ is special in two respects. First, there is a
massless pseudo-scalar in $\sfD_{[\pm2]}$, which has no analog for
$D>4$. It is given by
\bea |[\pm 2];-1;\pm 2,0\rangle&=& \e^{rst}M_{rs,-1}L^\pm_{t,1}|[\pm
2]\rangle\ .\label{psgs}\eea
Thus, in $D=4$
\bea \sfD_{[\pm 2]}&\supset & \sfD_{[\pm2];-1}(\pm
1,0)~\oplus~\sfD_{[\pm2];-1}(\pm 2,0)~\oplus~\sfD'_{[\pm
2];+}~\oplus~
\sfD'_{[\pm 2];-}\\[4pt]&\simeq&(\sfD^\pm_0)^{\otimes 2}\oplus
(\sfD^\pm_{\ft12})^{\otimes 2}\ ,\eea
where we have used \eq{FF1} and \eq{FFtheorem}. We note that the
massless pseudo-scalar is an $\widehat\msp(4)_{-1/2}$ descendant
of the massless scalar, while these spaces belong to distinct
irreps of the higher-spin algebra $\mhs(4)$, see {\emph
e.g.}~\cite{Konshtein:1988yg,Sezgin:2003pt}.

Second, the spinor singleton is hyperlight-like, as we saw in the
previous section. The affine extension of the spinor singleton
belongs to a $1$-twisted sector that is related by spectral flow
to the fundamental spinor of $\mso(2,3)$, \emph{i.e.}~the real
quartet of $\msp(4)$ which has lowest energy $-1/2$ and highest
energy $1/2$. More generally, the spectral flow gives rise to
spinorial $P$-twisted sectors. To describe these, we decompose the
quartet $SO(2)\times SO(3)$ into two $SO(3)$ doublets with energy
$\pm 1/2$, that we shall denote by
\bea |[0];\pm \rangle_i&=&|[0];\ft12;-\ft12,\ft12;\pm \rangle_i\
,\eea
where the quantum numbers indicate that
\bea L^\pm_{r,n}|[0];\pm \rangle_i&=& 0\
,\qquad n\geq 0\ ,\\[4pt]L^\mp_{r,n}|[0];\pm \rangle_i
&=&\pm\d_{n0}(\s_r)_i{}^j|[0];\pm
\rangle_j\ ,\qquad n\geq 0\ ,\\[4pt]
\left(E_n\mp \ft12\d_{n0}\right)|[0];\pm \rangle_i&=&0\ ,\qquad
n\geq 0\ ,\\[4pt] \left(M_{rs,n}|[0];\pm \rangle_i+ \ft{i}2
\d_{n0}(\s_{rs})_i{}^j|[0];\pm \rangle_j\right)&=& 0\ ,\qquad n\geq
0\ .\eea
The spectral flow operation yields the states
\bea |[P];\pm \rangle_i&=&
|[P];\ft{3-(|P|+1)^2}4;\ft{P+\s}{2},\ft12;\pm \rangle_i\ =\
\Omega_P(|[0];\pm \rangle_i)\ ,\eea
obeying
\bea L^\pm_{r,n}|[P];\pm \rangle_i&=& 0\ ,\qquad n\geq \pm P\ ,
\\[4pt]L^\mp_{r,n}|[0];\pm \rangle_i&=&\pm\d_{n\pm P,0}(\s_r)_i{}^j|[0];\pm
\rangle_j\ ,\qquad n\geq\mp P\ ,\qquad\\[4pt]
\left(E_n\mp\ft12\d_{n0}(|P|+1)\right)|[P];\pm \rangle_i&=&0\
,\qquad
n\geq 0\ ,\\[4pt] \left(M_{rs,n}|[0];\pm \rangle_i+ \ft{i}2
\d_{n0}(\s_{rs})_i{}^j\right)|[0];\pm \rangle_j&=& 0\ ,\qquad
n\geq
0\ ,\\[4pt]\left(L_n-\d_{n0}h_{[P];-\ft12,\ft12}\right)|[P];\pm
\rangle_i&=&0\ ,\qquad h_{[P];-\ft12,\ft12}\ =\ \ft{3-(|P|+1)^2}
4\ .\qquad\eea

The above analysis suggests that if one adds the above spinorial
sector to the model in $D=4$ (with multiplicities equal to $1$),
then the spinorial contributions to $\sfD_{[\pm 2]}$ make this
space isomorphic to $(\sfD^\pm_0\oplus \sfD^\pm_{1/2})^{\otimes 2}$.
The natural generalization would then lead to a
\bea \mbox{\it 4D extended model:}&&\sfD_{[P]}\ \simeq\
(\sfD^\sigma_0\oplus \sfD^\sigma_{\ft12})^{\otimes |P|}\
.\label{DPminus4}\eea

In Fig.~\ref{confdim} we summarize the set of ground states
specifying the representations $\sfD_{[P];h}(e_0,\mathbf m_0)$ in
the $D=4$ model for $|P|\le2$. 

\begin{figure}[t]
    \begin{center}
        \includegraphics[angle=-90,width=10cm]{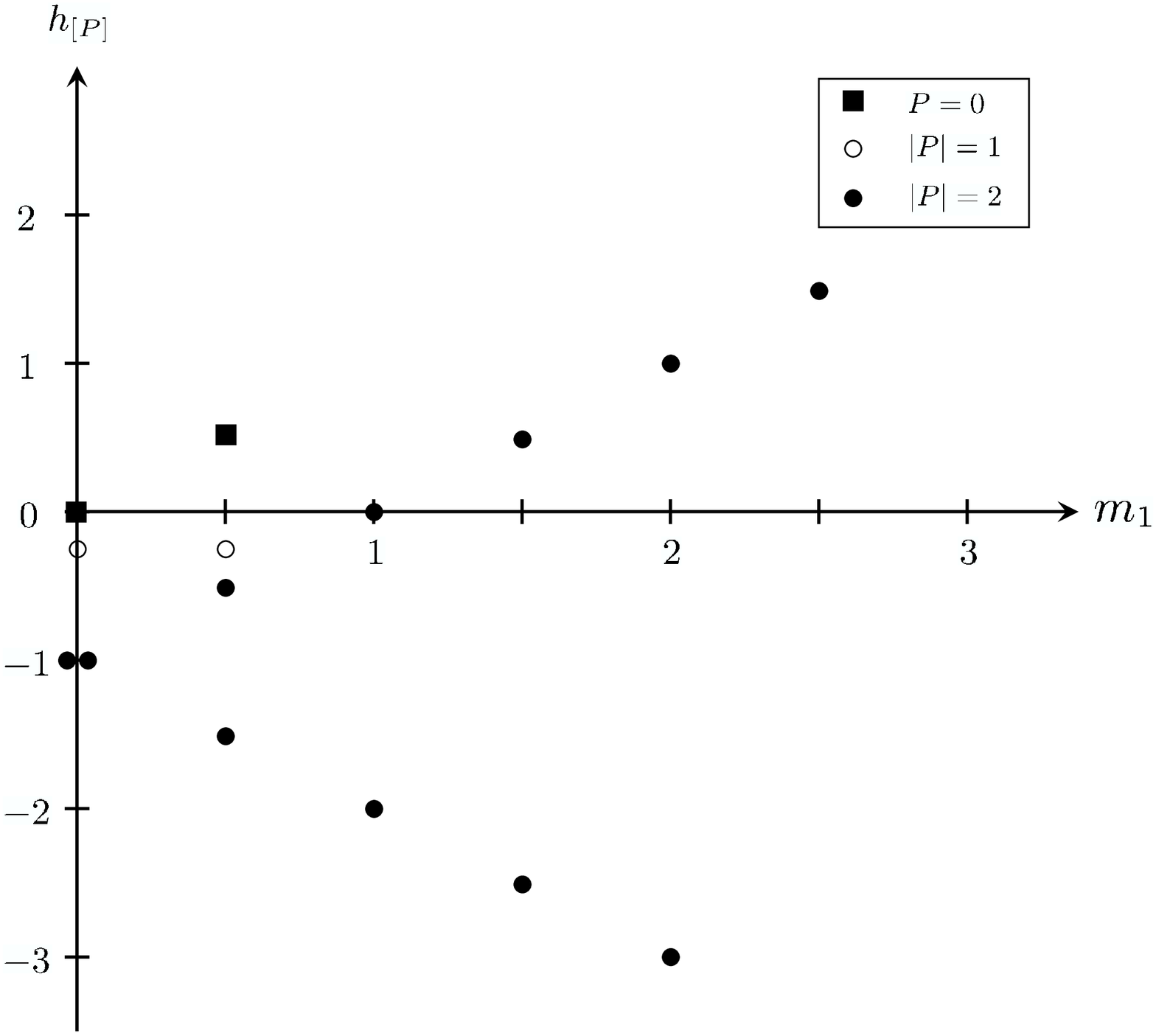}
    \end{center}
        \caption{The conformal dimensions $h_{[P]}$ as a function of the
spin $m_1=m$
        of the ground states of the $\mso(2,3)\simeq\msp(4)$ subspaces
$\sfD_{[P];h}(e_0,\mathbf m_0)$
        defined in \eq{DP}. Note that there are two scalar ground
        states having $h=-1$ in each of the $P=\pm2$ sectors, one of which
corresponds to
        the twisted primary state $|[\pm2];-1;\pm1,0\rangle$ and the other
to the pseudoscalar
        $|[\pm2];-1;\pm2,0\rangle$.}
    \la{confdim}
\end{figure}

Let us continue with a more detailed
analysis in $D=4$ using a realization in terms of symplectic bosons.

\scs{The Extended $\widehat{\msp}(4)$ Model}\label{Sec:4}


In this section we give the realization of the four-dimensional
extended model, which has symmetry group $\widehat
\mso(2,3)_{-1/2}\simeq \widehat\msp(4)_{-1/2}$, in terms of $4$ real
symplectic bosons forming a Majorana spinor
\cite{Feingold85,Goddard:1987td}\footnote{Also the models in $D=5$
and $D=7$ can be realized in terms of symplectic bosons in spinor
representations, but these models require additional internal
gaugings ($U(1)$ in $D=5$ and $SU(2)$ in $D=7$) that are critical
for scalar singletons but not for spinor (nor higher-spin)
singletons ground states. This provides an explanation why spinor
singletons are hyperlight-like in $D=4$ but not $D>4$.},
\emph{i.e.}~a quartet of $\msp(4)$. Indeed, the Sugawara central
charge $c=-2$ and conformal weights $h=-1/2$ and $h=-1/4$,
respectively, of the fundamental spinor and the twist field agree
with the canonical values of the system of symplectic bosons (where
the R sector corresponds to the twist field). To some extent, this
correspondence is analogous to that between the $\widehat\mso(4)_1$
model and $4$ real fermions. The important difference is that in the
fermionic model all conformal weights are bounded from below, so
that spectral flow becomes a $\integ_2$-operation, corresponding to
moving between weights in the two congruence classes of $SO(4)$, or
equivalently, between even and odd momenta in the two-dimensional
Euclidean lattice of the free-field realization. In the case of
symplectic bosons, the conformal weights are bounded from below by
$-P^2/4$ for fixed amount of twist, $P$. The spectral flow operation
now becomes a $\integ$-operation, corresponding to moving between
classes of momenta in the four-dimensional lattice of signature
$(2,2)$ of the free-field realization, which will be considered in
more detail in the next section.


\scss{Realization Using Symplectic Bosons}


In order to make contact with the standard oscillator realization
of the singletons, it is convenient to work in a $U(2)$-covariant
compact basis where the symplectic bosons are assembled into two
doublets $a_i(z)$ and $\bar a^i(z)$ ($i=1,2$),
obeying\footnote{Doublet indices are raised and lowered with
$\e^{ij}$ and $\e_{ij}$ defined so that $\e_{ik}\e^{jk}=\d_i^j$.
The symmetric Pauli matrices $(\s_r)_{ij}=(\s_r)_i{}^k\e_{kj}$ and
$(\s_r)^{ij}=\e^{ik}(\s_r)_{k}{}^j$ obey the reality condition
$\left((\s_r)_{ij}\right)^\ast=-(\s_r)^{ij}$.}
\bea \quad a_i(z)\bar a^j(w)\sim\frac{\d^j_i}{z-w}\
.\label{aope}\eea
In this basis, the $\widehat{\msp}(4)$ currents are given by
\bea \bar J^{ij}&=&\bar a^i\bar a^j\ , \qquad K^i{}_{j}\ =\ (\bar
a^i a_{j})\ , \qquad J_{ij}\ =\ a_{i}a_{j}\ , \eea
and obey
\bea \la{spfourcurrent1} J_{ij}(z)\bar
J^{kl}(w)&\sim&\frac{4\d_{(j}^{(l}K^{k)}{}_{i)}(w)}{z-w}-\frac{4k}{(z
-w)^2}\d^k_{(i}\d^l_{j)}\ , \\[4pt] \la{spfourcurrent2}
K^i{}_{j}(z)\bar J^{kl}(w)&\sim&\frac{\d_j^k\bar
J^{il}(w)}{z-w}+\frac{\d_j^l\bar J^{ik}(w)}{
z-w}\ ,\\[4pt] \la{spfourcurrent3}
K^i{}_{j}(z)J_{kl}(w)&\sim&-\frac{\d_k^iJ_{jl}(w)}{z-w}-\frac{\d_l^iJ_{jk}(w)}
{z-w} \ ,\\[4pt] \la{spfourcurrent4}
K^i{}_{j}(z)K^k{}_{l}(w)&\sim&\frac{\d_j^kK^i{}_{l}(w)-\d_l^iK^k{}_{j}(w
)}{z-w}+\frac{2k}{(z-w)^2}\d^i_l\d^k_j\ , \eea
where the normalization is chosen such that $k=-1/2$. The
corresponding commutation rules read
\bea \la{spfouraffine} [J_{ij,m},\bar
J^{kl}_{n}]&=&\d_j^lK^k{}_{i,m+n}+3~\mbox{terms}-4km\d^k_{(i}\d^l_{j
)}\d_{m+n,0}\ , \\[4pt][K^i{}_{j,m},\bar J^{kl}_n]&=&\d_j^k\bar
J^{il}_{m+n}+\d_j^l\bar J^{ik}_{m+n}\ ,\\[4pt]
[K^i{}_{j,m},J_{kl,n}]&=&-\d_k^iJ_{jl,m+n}-\d_l^iJ_{jk,m+n}\ ,\\[4pt]
[K^i{}_{j,m},K^k{}_{l,n}]&=&\d_j^kK^i{}_{l,m+n}-\d_l^iK^k{}_{j,m+n}+2km\d^i_l
\d^k_j\d_{m+n,0}\ . \eea
The $\widehat\msp(4)_{-1/2}$ Sugawara stress tensor equals the
canonical stress tensor of the symplectic bosons,
\bea T&=&\frac12 \left((a_i\del \bar a^i)-(\bar a^i\del
a_i)\right)\ \label{canT}\eea
with $c=-2$. The relation between $(\bar J^{ij},K^i{}_j,J_{ij})$
and the $O(3)$-covariant compact basis $(L^\pm_r,E,M_{rs})$ is
given by
\bea \bar J^{ij}&=& (\s^r)^{ij} L^+_r\ ,\qquad J_{ij}\ =\
-(\s^r)_{ij}L^-_r \ ,\\[4pt]K^i{}_j&=& \d^i_jE+\ft i2(\s^{rs})_j{}^i
M_{rs}\ .\eea
The currents $K^i{}_j$ generate the $\widehat\muu(2)_{-1/2}\simeq
\widehat \msu(2)_{-1/2}\oplus \widehat\muu(1)_{-1/2}\simeq
\widehat \mso(3)_{-1/2}\oplus \widehat\mso(2)_{-1/2}$ subalgebra.
There are also two independent $\widehat \msp(2)_{-1/2}\simeq
\widehat\mso(1,2)_{-1}$ subalgebras generated by the currents
($r'=+,3,-$)
\bea J^{r'(i)}&=&(J^{+(i)},J^{3(i)},J^{-(i)})\ =\ \ft12( \bar
J^{ii},K^i{}_i, J_{ii})\ ,\label{sp2curr}\eea
for fixed $i=1,2$. The two $\msp(2)$ spins are given in terms of
the space-time energy and spin by
\bea K^1{}_1&=&E+M_3\ ,\qquad K^2{}_2\ =\ E-M_{3}\ ,\eea
where we use the following canonical basis for $SO(3)$:
$M_r=-\e_{rst}M_{st}/2$, $[M_r,M_s]=i\epsilon_{rst}M_t$.
To represent the symplectic bosons in Fock spaces, one expands
them in modes
\bea \la{aexpansion}
a_i(z)&=&\sum_{n\in\integ+\mu}z^{-n-1/2}a_{i,n}\ , \qquad \bar
a^i(z)\ =\ \sum_{n\in\integ+\mu}z^{-n-1/2}\bar a^i_n\ , \eea
where by definition $\mu\in\{0,1/2\}$. The operator product
\eq{aope} is equivalent to the oscillator algebra
\bea [a_{m,i},\bar a^j_n]\ =\ \d_i^j\d_{m+n,0}\ .\eea
One next introduces {\it $P$-twisted oscillator vacua}
$|[P]\rangle$ obeying
\bea \la{ff1} a_{i,n}|[P]\rangle = 0\ ,~~ \mbox{for~
$n\ge-\ft{P-1}{2}$}\ ,\quad\quad \bar a^i_n|[P]\rangle=0\ ,~~
\mbox{for~ $n\ge\ft{P+1}{2}$}\ ,\eea
where $P\in 2\integ+2\mu$, giving rise to the $P$-twisted sectors
\bea \widehat {\cal F}_{[P]}&=& \left\{\prod_{n\geq
\ft{P+1}{2}}\left(a_{i,-n}\right)^{k_n} \prod_{n\geq
-\ft{P-1}{2}}\left(\bar a^i_{-n}\right)^{\bar
k_n}\vert[P]\rangle\right\}\ ,\eea
where $k_n,\bar k_n\in \integ_{\geq0}$. The $P$-twisted normal
order is defined by
\bea \ncross{\cal O}a_{i,n}\ncross&=& \mx{\{}{ll}{{\cal
O}a_{i,n}&\mbox{for~
$n\ge-\ft{P-1}{2}$}\\a_{i,n}{\cal O}&\mbox{else}}{.}\ ,\\[4pt]
\ncross{\cal O}\bar a^i_n\ncross&=&\mx{\{}{ll}{{\cal O}\bar
a^i_n&\mbox{for~ $n\ge\ft{P+1}{2}$}\\ \bar a^i_n{\cal
O}&\mbox{else}}{.}\ ,\eea
where we note that the odd-twisted normal orderings send
$a_{i,0}$ and $\bar a^i_0$ to the left and the right,
respectively. The $P$-twisted Green's functions are given by
\bea a_i(z)\bar a^j(w)-\ncross a_i(z)\bar a^j(w)\ncross
&=&\frac{\left({z\over w}\right)^{P/2}\d^j_i}{z-w}\ ,\eea
and the normal ordering only affects affects the energy operator
$E=(\bar a^i a_i)/2$ and the stress energy tensor $T=(\partial
\bar a^i a_i-\bar a^i\partial a_i)/2$. The $0$-twisted normal
order coincides with $(\cdot)$, and one can identify the
$0$-twisted sector with the standard NS sector, \emph{i.e.}
\bea |[0]\rangle&=&\vert 0\rangle\ ,\eea
is the $\widehat\msp(4)_{-1/2}$ invariant state
obeying $M_{AB,n}|0\rangle=L_n|0\rangle=0$ for $n\geq 0$. The
$(\pm 1)$-twisted sectors are analogs of the R-sector in the
realization of $\widehat\mso(4)_{1}$ in terms of $4$ real fermions
with $c_{\rm fermion}=2$ (here the currents are based on
anti-symmetric $SO(4)$-invariant bilinear forms; the central
extension in $\widehat\mso(4)_1$ is fixed by equating the Sugawara
and canonical stress tensors). In particular, the bosonic Fock
space generated by the zero-modes $(a_{i,0},\bar a^i_0)$ are
analogs of the finite-dimensional spinor representations of the
Clifford algebra with $h_{\rm spinor}=1/4$. Switching back to $4$
real bosons reverses the signs in $c$ and $h$ and induces a
non-trivial mixing between the stress tensor and the AdS energy
(whose fermionic analog vanishes identically due to the odd
statistics) with the result that
\bea E(z)&=&\ft12 \ncross\bar a^i a_i\ncross+\ft{P}{2}z^{-1}\
,\label{ET1}\\[4pt]
T(z)&=&\ft12\ncross(\partial \bar a^i a_i-\bar a^i\partial
a_i)\ncross-\ft{P}2\ncross\bar a^i a_i\ncross z^{-1}-\ft{P^2}{4}
z^{-2}\ ,\label{ET}\eea
where $P=\pm 1$. In fact, the above expression is valid for any
$P$, in agreement with \eq{flowE1} and \eq{flowT1}. One way of
showing this is to note that for $P=P'$ mod $2$ one can use simple
re-orderings of oscillators to go between the $P$-twisted and
$P'$-twisted normal ordered forms of operators corresponding to
states in the NS sector, \emph{i.e.}~monomials in the symplectic
bosons. The KM charges and Virasoro generators can consequently be
expanded in the $P$-twisted sector as follows
\bea \la{aa1} \bar J^{ij}_n&=&\sum_{m\in\integ+\mu}\ncross\bar
a^i_m \bar a^j_{n-m}\ncross\ , \qquad J_{ij,n}\ =\
\sum_{m\in\integ+\mu}\ncross a_{i,m}a_{j,n-m}\ncross\ ,
\\[4pt]
\la{aa2} K^i{}_{j,n}&=&\sum_{m\in\integ+\mu}\ncross\bar a^i_m
a_{j,n-m}\ncross+\ft{P}{2}\delta_{n,0}\d^i_j\ , \label{Kijmodes}\\[4pt]
L_n&=&{1\over 2}\sum_{m\in\integ+\mu}\ncross\left(m(\bar
a^i_{n-m}a_{i,m}- \bar a^i_{m}a_{i,n-m})-P\bar
a^i_{m}a_{i,n-m}\right)\ncross-\ft{P^2}4\d_{n,0}\ .
\label{Lnmodes}\eea

The star conjugation $*$ and BPZ conjugation $\tau$ defined in
\eq{conjgen} take the following form in the compact basis
\bea (J_{ij,n})^*&=&-\bar J^{ij}_{n}\ , \qquad (\bar J^{ij}_n)^*\
=\ -J_{ij,n}\
, \qquad (K^i{}_{j,n})^*\ = -K^j{}_{i,n}\ , \\[4pt]
(J_{ij,n})^\tau&=& -J_{ij,-n}\ , \quad (\bar J^{ij}_n)^\tau\ =\
-\bar J^{ij}_{-n}\ , \quad~~ (K^i{}_{j,n})^\tau\ =\ -K^i{}_{j,-n}\
,\\[4pt] (J_{ij,n})^\dagger
&=&\bar J^{ij}_{-n}\ ,\qquad (\bar J^{ij}_n)^\dagger\ =\
J_{ij,-n}\ ,\qquad (K^i{}_{j,n})^\dagger\ =\ K^j{}_{i,-n}\ .\eea
The conjugations can be implemented at the level of the oscillator
algebra by taking
\bea \la{oscconj1} (a_{i,n})^*&=& i\bar a^i_n\ , \qquad (\bar
a^i_{n})^*\ =\ i
a_{i,n}\ ,\\[4pt] \la{oscconj2}(a_{i,n})^\tau&=&-ia_{i,-n}\ , \qquad (\bar
a^i_{n})^\tau\ =\ -i\bar a^i_{-n}\ ,\\[4pt] (a_{i,n})^\dagger&=&\bar
a^i_{-n}\ ,\qquad (\bar a^i_{n})^\dagger\ =\ a_{i,-n}\ .
\la{oscconj3} \eea
The $\pi$ map, defined by \eq{pimapaffine1}, takes the following
form in the $U(2)$-covariant basis,
\bea \la{pimapaffine2} \pi(J_{ij,n})&=&\bar J_{ij,n}\ , \qquad
\pi(\bar J^{ij}_n)\ = \ J^{ij}_n\ , \qquad \pi(K^i{}_{j,n})\ = \
K_j{}^i\ =\ \e^{ik}K^l{}_k\e_{lj}\ ,\eea
and can be implemented as
\bea \pi(a_{i,n})\ = \ \bar a_{i,n}\ , \qquad \pi(\bar a^i_{n})\ =
\ a^i_n\ . \eea
Formally, the action of the conjugations and the $\pi$ map on
states can be defined as in \eq{conj1} and \eq{conj2}.


\scss{Generating the Spectrum by Spectral Flow}


To describe the $P$-twisted sectors by means of the spectral flow
operation, we define
\bea \vert[P]\rangle&=& \O_P\left[\vert 0\rangle\right]\ .\eea
From the fact that $\O_P[a_i(z)|0\rangle]$ and $\O_P[\bar
a^i(z)|0\rangle]$ are regular and non-vanishing at $z=0$ it
follows that
\bea (\O_P a_i\O_P^{-1})(z)&=& z^{-P/2} a_i(z)\ ,\qquad (\O_P \bar
a^i\O_P^{-1})(z)\ =\  z^{P/2} \bar a^i(z)\ ,\la{piPa}\eea
or, in terms of modes,
\bea \O_P a_{i,n} \O_P^{-1}&=&a_{i,n-P/2}\ ,\qquad \O_P \bar
a^i_{n} \O_P^{-1}\ =\ \bar a^i_{n+P/2}\ .\eea
The spectral flows of $E(z)$ and $T(z)$ can then be calculated
either by implementing the normal order using an auxiliary
integration and following steps similar to those that led from
\eq{OPsugawara} to \eq{flowT}, or by acting directly on the mode
expansions in \eq{Kijmodes} and \eq{Lnmodes}. The result can be
written as
\bea \O_P E(z)\O_P^{-1}&=& E(z) -\ft{P}2 z^{-1}\ ,
\label{tildeE2}\\[4pt]
\O_P T(z)\O_P^{-1}&=&T(z) + \ft{P}{z}E(z)- \ft{P^2}4 z^{-2} \ ,
\label{tildeT2}\eea
where the left-hand sides above are in $P$-twisted normal order,
which ensures agreement with \eq{ET1} and \eq{ET}. It is now
straightforward to verify that
\bea \la{xx1}\bar J^{ij}_n\vert[P]\rangle&=& 0\ ,\qquad n\geq P\ ,\\[4pt]\la{xx2}
J_{ij,n}\vert[P]\rangle&=& 0\ ,\qquad n\geq -P\ ,\\[4pt]\la{xx3}
\left(K^i{}_{j,n}-\ft{P}2
\delta_{n,0}\delta^i_j\right)\vert[P]\rangle
&=& 0\ , \qquad n\geq 0\ ,\\[4pt]\la{xx4}
\left(L_n+\ft{P^2}4\delta_{n,0}\right)\vert[P]\rangle&=& 0\ ,
\qquad n\geq 0\ ,\eea
\emph{i.e.}~$\vert[P]\rangle$ corresponds to the
$\mso(2,3)_{[P]}$-singlet $P$-twisted primary defined in \eq{twpr1}
and \eq{twpr2}. In addition to the singlet, there is a twisted
primary spinor, given for $P\neq 0$ by
\bea |[P];-\ft{(|P|+1)^2}4+\ft34;\ft{(P+\sigma)}2,\ft12)\rangle&=&
\left\{\ba{ll}
\bar a^i_{(P-1)/2}|[P]\rangle&P>0\\[8pt]
a^i_{-(P+1)/2}|[P]\rangle &P<0\ea\right.\ ,\eea
and for $P=0$ by the quartet (denoted here as a lowest weight
state)
\bea |[0];\ft12;-\ft 12,\ft12\rangle&=& \left\{\bar
a^i_{-\ft12}|0\rangle\ ,~a^i_{-\ft12}|0\rangle\right\}\ .\eea
As shown in Appendix A, there are no other primary states, so that
\bea \widehat {\cal F}_{[P]}&=& \widehat \sfD_{[0];0}(0,0)\oplus
\widehat \sfD_{[0];\ft12}(-\ft12,\ft12)\nn\\[4pt]&&\bigoplus_{P\neq 0}
\left[\widehat \sfD_{[P];-\ft{P^2}4}(\ft{P}2,0)\oplus \widehat
\sfD_{[P];-\ft{(|P|+1)^2}4+\ft34}(\ft{(P+\sigma)}2,\ft12)\right]\
.\label{FPdec}\eea
%


\scss{Oscillator realization of the Multipleton Subsectors}


Let us make contact with Dirac's original oscillator realization
of the $\msp(4)$ singletons. For $P\neq 0$ we split the
oscillators $(\bar a^i_{n},a_{i,n})$ ($n\in \integ +P/2$) into
$P$-twisted non-zero modes with $|n|\geq (|P|+1)/2$ and zero modes
with $|n|\leq (|P|-1)/2$. The zero modes can be assembled into
$|P|$ sets of $U(2)$-covariant oscillators $(b_i(\xi),\bar
b^i(\xi))$, $\xi=1,\dots,|P|$, obeying
\bea [b_i(\x),\bar b^j(\eta)]&=&\delta_{\xi\eta}\delta_i^j\ ,\eea
\bea P>0&:& b_i(\xi)|[P]\rangle\ =\ 0\ ,\\[4pt]
P<0&:& \bar b^i(\xi)|[P]\rangle\ =\ 0\ .\eea
This can be described geometrically in the complex plane as the
conformal transformation $z=\exp(\zeta/P)$, which maps the
symplectic bosons to
\bea b_i(\zeta)&\equiv & \left({dz\over d\zeta}\right)^{1/2}
a_i(z(\zeta))\ =\ {1\over
\sqrt{P}} \sum_{n} e^{-{n\zeta\over P}} a_{i,n}\ ,\\[4pt]
\bar b^{i}(\zeta)&\equiv & \left({dz\over d\zeta}\right)^{1/2}
\bar a^i(z(\zeta)) \ =\ {1\over \sqrt{P}} \sum_{n}
e^{-{n\zeta\over P}} \bar a^i_n\ ,\eea
followed by a truncation to zero modes, defined by
\bea b_i^{(P)}(\zeta)\ = {1\over \sqrt{P}} \sum_{|n|\le (|P|-1)/2}
e^{-{n\zeta\over P}} a_{i,n}\ ,\qquad \bar b^{(P)i}(\zeta)\ =
{1\over \sqrt{P}}\sum_{|n|\le (|P|-1)/2} e^{-{\zeta\over P}} \bar
a^i_n\ ,\eea
after which one identifies
\bea b_i(\xi)&=& b^{(P)}_i(2\pi i \x)\ ,\qquad \bar b^i(\xi)\ =\
\bar b^{(P)i}(2\pi i \x)\ .\eea
The zero modes yields a subspace of ${\cal
F}_{[P]}\subset\widehat{\cal F}_{[P]}$ given by
\bea {\cal F}_{[P]} &=& \mx{\{} {ll} {\left\{
\prod_{\xi,i} \bar b^i(\x) |[P]\rangle\right\}&\mbox{for $P>0$}\ ,\\[10pt]
\left\{ \prod_{\xi,i}b_i(\x)|[P]\rangle\right\}&\mbox{for
$P<0$}}{.}\nn\\[4pt] &\simeq &  \left({\cal F}^{{\rm
sign}(P)}\right)^{\otimes |P|}\ ,\qquad P\neq 0\
,\label{calFP1}\eea
where ${\cal F}^{+}$ denotes the standard Fock space and ${\cal
F}^{-}$ the anti-Fock space of a $U(2)$-covariant oscillator
$(b_i,\bar b^i)$ obeying $[b_i,\bar b^j]=\d_i^j$. The latter is
obtained by acting with $b_i$ on an anti-vacuum $\vert 0\rangle^-$
obeying $\bar b^i\vert 0\rangle^-=0$. The algebra $\msp(4)$ is
represented in ${\cal F}^{\pm}$ by
\bea \bar J^{(b)ij}&=&\bar b^i\bar b^j\ ,\qquad K^{(b)i}{}_j\ =\
\bar b^i b_j +\ft12 \delta^i_j\ =\ b_j\bar b^i-\ft12\d^i_j\
,\qquad J^{(b)}_{ij}\ =\ b_ib_j\ ,\label{sp4osc}\eea
leading to the decompositions
\bea {\cal F}^\pm&=& \sfD_0^\pm\oplus \sfD_{1/2}^\pm\
,\label{calF}\eea
where the scalar and spinor singletons and anti-singletons are
realized as ($m=0,1/2$)
\bea \sfD_{m}^+=\big\{\bar b^{i_1}\cdots \bar
b^{i_{2(n+m)}}|[1]\rangle\big\}_{n=0}^\infty\ ,\qquad
\sfD_{m}^-=\big\{b_{i_1}\cdots
b_{i_{2(n+m)}}|[-1]\rangle\big\}_{n=0}^\infty \
.\label{singletonosc}\eea
The ground states $|1/2,0\rangle^\pm$ and $|1,1/2\rangle^{+}_i$
and $|1,1/2\rangle^{-}_i$ obey
\bea \la{singcond1} \mbox{Scalar:} &&
K^{(b)i}{}_j|\ft12,0\rangle^\pm\ =\ \pm
\ft12\d_j^i|\ft12,0\rangle^\pm\ ,\\[4pt]&& J^{(b)}_{ij}|\ft12,0\rangle^+\
=\ 0\ ,\qquad
\bar J^{(b)ij}|\ft12,0\rangle^-\ =\ 0\ , \\[10pt]
\la{singcond2} \mbox{Spinor:} &&
K^{(b)i}{}_j|1,\ft12\rangle^{+}_k\ =\
\pm\left(\ft12\d_j^i|1,\ft12\rangle^{+}_k+\e_{jk}|1,\ft12\rangle^{+i}\right)\
,\\[4pt] && J^{(b)}_{ij}|1,\ft12\rangle^{+}_k\ =\ 0\
,\qquad \bar J^{(b)ij}|1,\ft12\rangle^{-}_k\ =\ 0\
.\qquad\qquad\eea
Thus, for $P\neq 0$ the space ${\cal F}_{[P]}$ consists of all
possible $P$-tupletons built from scalar as well as spinor
(anti-)singletons of $\msp(4)$ (we note that the permutation group
$\cS_P$, which acts on $\xi$, and the Virasoro generator $L_0$ do
not commute). For $P=0$ we instead define
\bea \la{dec1zero} {\cal F}_{[0]}&=& \sfD_{[0];0}(0,0)\oplus
\sfD_{[0];\ft12}(-\ft12,\ft12)\ ,\eea
which form a finite-dimensional reducible $\mso(2,3)$ representation
space.

Let us examine the cases $P=\pm1$ and $P=\pm2$ in more detail. In
$\widehat {\cal F}_{[\pm 1]}$ we find the subspace ${\cal F}_{[\pm
1]}$ generated by the zero modes $(b_i(1),\bar b^i(1))=(a_{i,0},
\bar a^i_0)$. Here, the KM charges $(\bar
J^{ij}_n,K^i{}_{j,n},J_{ij,n})$ vanish for $n>0$ and reduce to the
oscillator representation \eq{sp4osc} for $n=0$, leading to an
oscillator realization of the $\msp(4)$ singletons and
anti-singletons of the form given in \eq{singletonosc}. For $P=\pm
2$, the zero modes
\bea b_i(1)&=&{i\over \sqrt{2}} (a_{i,-1/2}-a_{i,1/2})\ ,\qquad
\bar b^i(1)\ =\ {i\over \sqrt{2}}
(\bar a^i_{-1/2}-\bar a^i_{1/2})\ ,\\[4pt] b_i(2)&=&{1\over
\sqrt2}(a_{i,-1/2}+a_{i,1/2})\ ,\qquad \bar b^i(2)\ =\ {1\over
\sqrt2}(\bar a^i_{-1/2}+\bar a^i_{1/2})\ ,\eea
generate the subspaces ${\cal F}_{[\pm 2]}\subset \widehat{\cal
F}_{[\pm 2]}$, that decompose into massless $\msp(4)$
representations in accordance with \eq{FFtheorem}. Diagonalizing
$L_0$, one finds that
\bea {\cal F}_{[\pm 2]}&\simeq& (\sfD^\pm_0\oplus
\sfD^\pm_{1/2})^{\otimes 2}\ ,\eea
in agreement with \eq{calFP1} and \eq{calF}. The ground states are
given by
\bea |[2];\pm m-1;m+1,m\rangle^{i(2m)}&=&\bar
a_{\mp\ft12}^{i_1}\cdots
\bar a_{\mp\ft12}^{i_{2m}}|[2]\rangle\ , \\
|[-2];\pm m-1;-m-1,m\rangle_{i(2m)}&=&a_{i_1,\mp\ft12}\cdots
a_{i_{2m},\mp\ft12}|[-2]\rangle\ , \\
|[2];-1;2,0\rangle&=&\e_{ij}\bar a_{\ft12}^{i} \bar
a_{-\ft12}^{j}|[2]\rangle\ ,
\\ |[-2];-1;-2,0\rangle&=&\e^{ij}a_{i,\ft12} a_{j,-\ft12}|[-2]\rangle\ . \eea
We note that all ground states are built from oscillators with the
same Virasoro mode numbers except the massless pseudo-scalar
ground state, which we can also write as
$|[2];-1;2,0\rangle=\e_{ij}\bar J^{ki}_1
K^{j}{}_{k,-1}|[2];-1;1,0\rangle$ in agreement with \eq{psgs}. The
KM charges $(\bar J^{ij}_n,K^i{}_{j,n},J_{ij,n})$ with $n>1$
vanish in ${\cal F}_{[\pm 2]}$. The ground states in ${\cal
F}_{[2];-}$ can be written as strings of $\bar J^{ij}_1$ charges
acting on the scalar ground state $|[2]\rangle$ and the spinor
ground state $\bar a^i_{-1/2}|[2]\rangle$. On the other hand,
$\bar J^{ij}_{-1}$ yields the massless higher-spin ground states
in ${\cal F}_{[2];+}$ together with extra contributions involving
oscillator non-zero modes that can be cancelled by adding terms
involving $K^i{}_{j,-1}$ of the form given in \eq{photon}.

So far, the realization of the $\widehat\mso(2,3)_{-1/2}$ in terms
of symplectic bosons supports the claim made in \eq{DPminus4},
with the identification
\bea \sfD_{[P]}&=&{\cal F}_{[P]}\ .\eea
Next, we wish to continue and analyze in more detail the fusion
rules and locality properties of the operator products using a
free-field realization.


\scs{Free-Field Realization of the Extended $\widehat{\msp}(4)$ Model}\la{Sec:5}


In this section we use a free-field realization to analyze the
fusion rules and locality properties of the operator algebra of
the extended $\widehat{\msp}(4)$ model introduced in Section
\ref{Sec:4}. This realization also provides a very simple
implementation of the spectral flow $\O_P$.


\scss{Partial Bosonization}


To construct the $\widehat\msp(4)_{-1/2}$ model out of free fields
one in principle needs to introduce four real free bosons,
$(\varphi^i,\sigma^i)$, $i=1,2$, in a space with signature $(--++)$
and background charges $(0,0,i/\sqrt{2},i/\sqrt{2})$. However, when
it comes to describing the states of the Fock spaces introduced in
the previous section, it is more practical to work in a partially
bosonized picture analogous to that of the $\widehat\msp(2)_{-1/2}$
model worked out in \cite{Lesage:2002ch} (see also
\cite{Bouwknegt:1988iz}). Here, one retains the two time-like
components $\varphi^i$ ($i=1,2)$, defined by
\bea \label{opedefphi} \varphi^i(z)\varphi^j(w)&\sim&
\d^{ij}\ln(z-w)\ ,\qquad T_\varphi\ =\
\ft12(\del\varphi^i\del\varphi^i)\ ,\eea
and replaces the two space-like bosons by two sets of Grassmann
odd weight $(0,1)$ fields $(\xi^i,\eta^i)$ ($i=1,2$), defined by
\bea \label{opexieta}\xi^i(z)\eta^j(w)&\sim& -\eta^j(w)\xi^i(z)\
\sim\ \frac{\d^{ij}}{z-w}\ , \qquad T_{(\x,\e)}\ =\
(\del\xi^i\eta^i)\ .\eea
Giving these fields the following mode expansions
\bea \varphi^i(z)&=&q^i-ip^i\ln
z+i\sum_{n\neq0}\frac{\a^i_{n}}{n}z^{-n}\ ,\eea\\[-25pt]\bea \xi^i(z)&=&
\sum_{n\in\integ} \xi^i_nz^{-n}\ , \qquad \eta^i(z)\ =\
\sum_{n\in\integ} \eta^i_{n}z^{-n-1}\ , \eea
the operator product expansions correspond to the commutation
rules
\bea [q^i,p^j]&=&-i\d^{ij}\ , \qquad [\a^i_{m},\a^j_{n}] \ =\
-m\d^{ij}\d_{m+n,0}\ ,\qquad \{\xi^i_m,\eta^j_{n}\}\ =\
\d^{ij}\d_{m+n,0}\ ,\qquad\eea
and the NS vacuum is the state obeying
\bea \a^i_{n}|0\rangle\ =\
\xi^i_{n+1}|0\rangle&=&\eta^i_{n}|0\rangle\ =\ 0\qquad \mbox{for}~
n\ge0\ , \eea
where $\a^i_{0}\equiv p^i$, so that $\varphi^i(z)|0\rangle$,
$\xi^i(z)|0\rangle$ and $\eta^i(z)|0\rangle$ are regular and
non-vanishing at $z=0$. For operators that do not depend on $q^i$,
\emph{i.e.} ${\cal O}(\del\varphi^i,\xi^i,\eta^i)$, the normal
ordering $:{\cal O}:$ is defined via the standard prescription
$(\cdot)$, which is an associative and commutative operation for
free fields. The ordering of operators that also depend on $q^i$
is defined by declaring that $:q^i{\cal O}:=q^i:{\cal O}:$. The
total Virasoro generators are given by
\bea
L_n&=&-\ft12\sum_{m\in\integ}\a^i_{n-m}\a^i_{m}-\sum_{m\in\integ}
m\xi^i_{m}\eta^i_{n-m}\ ,\qquad n\neq 0\ , \\[4pt]
L_0&=&-\ft12p^i
p^i-\sum_{m>0}\a^i_{-m}\a^i_{m}+\sum_{m>0}m\left(\xi^i_{-m}\eta^i_{m}+\eta^i_{-m}\xi^i_m\right)\
, \eea
and the total central charge $c=2\times 1+2\times(-2)=-2$. The
vertex operators
\bea V_{\lambda}&=&:c_\l(p) \exp(-i\lambda^i\varphi^i):\ ,\qquad
c_\l(p)\ =\ \exp\left({i\pi\over 2}\e^{ij}\l^ip^j\right)\eea
have conformal weights $h_\lambda=-\ft12\lambda^2$, where
$\lambda^2=\lambda^i\lambda^i$, and create states carrying finite
momenta,
\bea |\l\rangle\ = \ e^{-i\l^i q^i}|0\rangle\ , \qquad
(p^i-\l^i)|\l\rangle\ = \ 0\ .\eea
The vertex operators obey the composition rule
\bea V_\l(z)V_{\l'}(w)&=& e^{{i\pi\over 2}\e^{ij}\l^i\l^{\prime
j}}(z-w)^{-\l\l'}:c_{\l+\l'}(p)V_\l(z) V_{\l'}(w):\ .\eea
Reversing the order of the product and continuing analytically
using $w-z=e^{\pm i\pi}(z-w)$ yields the following monodromy
matrices,
\bea V_\l(z)V_{\l'}(w)&=& M^{\pm}_{\l,\l'}V_{\l'}(w)V_\l(z)\
,\\[4pt]M^\pm_{\l,\l'}&=&
\exp\left(i\pi(\e^{ij}\pm\d^{ij})\l^i\l^{\prime j}\right)\
.\label{monodromy1}\eea
The realization of the symplectic bosons now takes the form
\bea a_i&=&V_{\l_{(i)}}\eta^i\ =\
:c_{\l_{(i)}}(p) e^{-i\l_{(i)}^j\varphi^j}:\,\eta^i\ ,\label{bosonizeda1}\\[4pt]\bar a^i&=&
V_{-\l_{(i)}}\del\xi^i\ =\ :c_{-\l_{(i)}}(p)
e^{i\l_{(i)}^j\varphi^j}:\,\del\xi^i\ ,\label{bosonizeda2}\eea
where $\l_{(1)}=(1,0)$ and $\l_{(2)}=(0,1)$. The expansions
\eq{aexpansion} correspond to momenta in the lattice
\bea \l^i\in \integ+\mu\ ,\qquad i=1,2\ .\eea
The $\ast$ map \eq{oscconj1}, which can be written as $(a_i(\bar
z))^\ast=i\bar a^i(z)$ and $(\bar a^i(\bar z))^\ast=ia_i(z)$,
requires
\bea (\varphi^i(\bar z))^\ast&=&\varphi^i(z)\ ,\qquad (\eta^i(\bar
z))^\ast\ =\ i\del\xi^i(z)\ .\eea
The latter map can be implemented in the space of \emph{reduced
states} obeying
\bea \eta^i_{0}|\Psi\rangle&=&0\ .\eea
Here, the $\ast$ map and conjugations take the form\footnote{We
define $(AB)^\tau = (-1)^{AB} B^\tau A^\tau$ and $(AB)^\dagger =
(-1)^{AB} B^\dagger A^\dagger$.}
\bea (q^i)^*&=& q^i\ ,\qquad (\a^i_{n})^*\ =\ -\a^i_{n}\ ,\qquad
(\eta^i_{n})^*\ =\ -in\xi^i_n\ ,\qquad (\xi^i_n)^\ast\ =\ -{i\over
n}
\eta^i_{n}\ ,\\[4pt]
(q^i)^\t &=& q^i\ ,\qquad (\a^i_{n})^\t = -\a^i_{-n}\ ,\qquad
(\eta^i_{n})^\t\ =\ -\eta^i_{-n}\ , \qquad\!\!\!
(\xi^i_{n})^\t\ =\ \xi^i_{-n}\\[4pt]
(q^i)^\dagger&=&q^i\ ,\qquad (\a^i_{n})^\dagger\ =\ \a^i_{-n}\
,\qquad (\eta^i_{n})^\dagger\ =\ in\xi^i_{-n}\ ,\qquad
(\xi^i_n)^\dagger\ =\ -{i\over n}\eta^i_{-n}\ .\qquad\qquad\eea
where $n\neq 0$ for the fermionic oscillators. In the realization
\eq{bosonizeda1}, the $\widehat\msp(4)_{-1/2}$ currents read
(\emph{cf.}~\eq{spfourcurrent1}--\eq{spfourcurrent4})
\bea J_{ij}&=&V_{\l_{(i)}+\l_{(j)}}\chi_{ij}\ =\
:c_{\l_{(i)}+\l_{(j)}}(p)e^{-i(\varphi^i+\varphi^j)}:\chi_{ij}\ ,
\\[4pt]\bar J^{ij}&=&V_{-\l_{(i)}-\l_{(j)}}\bar \chi^{ij}\ =\
:c_{-\l_{(i)}-\l_{(j)}}(p) e^{i(\varphi^i+\varphi^j)}:\bar
\chi^{ij}\ ,\\[4pt] K^i{}_j&=&
V_{\l_{(i)}-\l_{(j)}}\chi^i{}_j\ =\
:c_{\l_{(i)}-\l_{(j)}}(p)e^{i(\varphi^j-\varphi^i)}:\chi^i{}_j\ ,
\eea
where
\bea \nn \chi_{ij}&=&\left\{\begin{array}{l}
  \del\eta^{i}\eta^{j}\ , \qquad i=j \\[4pt]
  \eta^{i}\eta^{j}\ , \qquad~ i\neq j
\end{array}\right.\ ,  \qquad  \bar \chi^{ij}\ =\ \left\{\begin{array}{l}
  \del^2\xi^i\del\xi^j\ , \qquad i=j \\[4pt]
  \del\xi^{i}\del\xi^{j}\ , \qquad ~i\neq j
\end{array}\right.\ ,  \\[4pt] \chi^i{}_j&=&\left\{\begin{array}{l}
  -i\del\varphi^i\ , \qquad i=j \\[4pt]
  \del\xi^2\eta^1\ , \qquad i=2,~ j=1 \\[4pt] \del\xi^1\eta^2\ ,\qquad
  i=1, ~j=2
\end{array}\right.\ .
\eea
We note that while the cocycle factors $c_\l(p)$ play a
non-trivial role in reproducing the correct monodromy properties
of the symplectic bosons \eq{bosonizeda1} and \eq{bosonizeda2}, they
do not contribute phase factors to the operator product between
the currents. Next, we shall turn to the bosonization of the twist
fields.


\scss{Singleton Twist Fields and Their Fusion Rules}


The singleton and anti-singleton states in ${\cal F}_{[\pm1]}$
correspond to a set of twist fields
\bea \S_{[\pm1]}^{(e,m)}&=&\left\{
\S_{[\pm1];-1/4}^{(e,m;\ell)}\right\}_{\ell=-m}^{m}\ ,\qquad e\ =\
\pm(m+\ft12)\ ,\eea
where $m\in\integ_{\geq0}$ and $m\in\integ_{\geq0}+1/2$ for scalar
and spinor singletons, respectively, that generate anti-periodic
branch cuts in the symplectic bosons in accordance with \eq{ff1}
for $P=\pm1$. This requirement determines the twist fields, and
one finds ({\it cf.}~\cite{Lesage:2002ch})
\bea \S^{(\pm
m\pm1/2,m;\ell)}_{[\pm1];-1/4}&=&V_{\mp(m+\ell+\ft12,m-\ell+\ft12)}\s^{(m;\ell)}_{\pm}\
, \eea
where
\bea \la{sigmadef1}
\s^{(m;\ell)}_{+}&=&(\del\xi^1\del^2\xi^1\cdots\del^{m+\ell}\xi^1)
(\del\xi^2\del^2\xi^2
\cdots\del^{m-\ell}\xi^2)\ , \\[4pt]
\la{sigmadef2}
\s^{(m;\ell)}_{-}&=&(\eta^1\del\eta^1\cdots\del^{m+\ell-1}\eta^1)
(\eta^2\del\eta^2 \cdots\del^{m-\ell-1}\eta^2)\ , \eea
with the convention that $\partial\xi^1\cdots
\partial^0\x^1\equiv 1$ idem $\xi^2$, $\eta^1$ and $\eta^2$. Indeed,
the conformal weights $h\left[\S_{[\pm1]}^{(e,m)}\right]=-\ft14$,
since $h\left[\s_{\pm}^{(m;\ell)}\right]= m(m+1)+\ell^2$ and
$h\left[e^{i(m\pm \ell+1/2)\varphi^i}\right]=-\ft{1}{2}(m\pm
\ell+\ft12)^2$. In particular, the products
$J_{ij}(z)\S^{(\pm1/2,0)}_{[1];-1/4}(w)$, $\bar
J^{ij}(z)\S^{(\pm1/2,0)}_{[-1];-1/4}(w)$ and
$M_3(z)\S^{(\pm1/2,0)}_{[1];-1/4}(w)$, where
$M_3=(K^1{}_1-K^2{}_2)/2$, have trivial expansions, while
\bea E(z)\S^{(\pm1/2,0)}_{[\pm1];-1/4}(w)&\sim&\frac{\pm1/2}{z-w}
\S_{[\pm1];-1/4}^{(\pm1/2,0)}(w)\ ,\eea
where $E=(K^1{}_1+K^2{}_2)/2$. Similarly,
$\S_{[\pm1];-1/4}^{(\pm1,1/2;\ell)}$ ($\ell=\pm1/2$) correspond to
the spinor ground states. We note that the anti-periodicity of the
branch cut, \emph{i.e.}
\bea a_i(e^{2\pi i}z)\S_{[\pm1]}^{(e,m)}(w)&=&
-a_i(z)\S_{[\pm1]}^{(e,m)}(w)\ ,\\[4pt]a^i(e^{2\pi
i}z)\S_{[\pm1]}^{(e,m)}(w)&=&-a^i(z)\S_{[\pm1]}^{(e,m)}(w)\ ,\eea
for $|z|>|w|$, does not rely on the cocycle factor.

Turning to the fusion rules in the $|P\leq 1$ sector, some of the
simplest cases are
\bea \nn \S_{[1];-1/4}^{(1/2,0)}(z)\S_{[-1];-1/4}^{(-1/2,0)}(w)
&\sim& (z-w)^{1/2}\ , \\[4pt]
\S^{(1/2,0)}_{[1];-1/4}(z)\S^{(-3/2,1;1)}_{[-1];-1/4}(w)
&\sim&(z-w)^{3/2}J_{11}(w)\ , \eea
where the cocycle factors have been suppressed. More generally,
for $m+\ell\ge m'+\ell'$ and $m-\ell\ge m'-\ell'$ one finds that
\bea
\S^{(e,m;\ell)}_{[1];-1/4}(z)\S^{(e,m';\ell')}_{[-1];-1/4}(w)\
\propto \ (z-w)^{m-m'+1/2}:(\bar a^1)^{m+\ell-m'-\ell'}(\bar
a^2)^{m-\ell-m'+\ell'}:(w)\ ,\label{fus1}\eea
where the right-hand sides are descendants in the NS sector that
can be written in terms of undifferentiated currents. Related
expansions hold for other relations among $(m,\ell,m',\ell')$.
Similarly, the products of twist fields from ${\widehat{\mathfrak
D}_{[1];-1/4}(1/2,0)}$ and and ${\widehat{\mathfrak
D}}_{[0];1/2}(-1/2,1/2)$ belong to ${\widehat{\mathfrak
D}_{[1];-1/4}(1,1/2)}$; for example
\bea \bar a^1(z)\S_{[1];-1/4}^{(1/2,0)}(w)\ \sim \
\frac{\S_{[1];-1/4}^{(1,1/2;1/2)}(w)}{(z-w)^{1/2}}\ , \qquad \bar
a^2(z)\S_{[1];-1/4}^{(1/2,0)}(w)\ \sim \
\frac{\S_{[1];-1/4}^{(1,1/2;-1/2)}(w)}{(z-w)^{1/2}}\ .\eea
Proceeding in this fashion, it is straightforward to verify the
following fusion rules within the $|P|\leq 1$ sector:
\bea  {\widehat{\mathfrak D}_{[\pm1]}(\pm\ft12,0)}\times
{\widehat{\mathfrak D}_{[\mp1]}(\mp\ft12,0)}&=&\sfDid\ , \\[4pt]
 {\widehat{\mathfrak D}_{[\pm1]}(\pm1,\ft12)}\times
{\widehat{\mathfrak D}_{[\mp1]}(\mp1,\ft12)}&=&\sfDid\ , \\[4pt]
{\widehat{\mathfrak D}_{[\pm1]}(\pm\ft12,0)}\times
{\widehat{\mathfrak D}_{[\mp1]}(\mp1,\ft12)}&=&\sfDq\ ,\\[4pt]
{\widehat{\mathfrak D}_{[\pm1]}(\pm\ft12,0)}\times \sfDq &=&
{\widehat{\mathfrak D}_{[\pm1]}(\pm1,\ft12)}\  , \\[4pt]
{\widehat{\mathfrak D}_{[\pm1]}(\pm1,\ft12)}\times \sfDq &=&
{\widehat{\mathfrak D}_{[\pm1]}(\pm\ft12,0)}\  , \\[4pt]  \sfDq\times
\sfDq&=&  \sfDid\ , \eea
where the conformal dimensions are suppressed. Next, let us
examine how the product of two singleton twist fields closes on a
massless twist field.


\scss{Massless Sector}

The formal analysis in Section \ref{Sec:3} implies that the product
of two singleton twist fields in the $P=1$ sector closes on a twist
field in the $P=2$ sector, forming an affine version of the
Flato-Fr\o  nsdal formulae \eq{FF1}-\eq{FFtheorem}. As an example,
let us consider the product of two singleton ground-state twist
fields:
\bea \S^{(1/2,0)}_{[1];-1/4}(z)\S^{(1/2,0)}_{[1];-1/4}(w)&\sim&
\frac{1}{(z-w)^{1/2}}\S^{(1,0)}_{[2];-1}(w) \ .\eea
As the notation indicates, the field
\bea \S^{(1,0)}_{[2];-1}&\equiv& \S^{(1,0;0)}_{[2];-1}\ =\
V_{-(1,1)}\ =\ :c_{-(1,1)}e^{i(\varphi^1+\varphi^2)}:\ ,\eea
is the ground state of the affine massless scalar representation
${\widehat{\mathfrak D}_{[2];-1}(1,0)}$, obeying
\bea && E(z)\S^{(1,0)}_{[2];-1}(w)\ \sim \
\frac{1}{z-w}\S^{(1,0)}_{[2];-1}(w)\ ,
\qquad M_3(z)\S^{(1,0)}_{[2];-1}(w)\ \sim\ 0\ , \\
&& J_{ij,n}\S^{(1,0)}_{[2];-1} \ =\ 0\quad \mbox{for}~n\ge-2\ ,
\qquad
 \bar J^{ij}_{n}\S^{(1,0)}_{[2];-1} \ =\ 0 \quad
~\mbox{for}~n\ge2\ , \eea
in accordance with \eq{xx1} -- \eq{xx4}. Another illustrative
example is
\bea \S^{(1/2,0)}_{[1];-1/4}(z)\S^{(3/2,1;0)}_{[1];-1/4}(w)&\sim&
\frac{\S^{(2,1;0)}_{[2];-2}(w)}{(z-w)^{3/2}}
 +{\ft{1}4:c_{-(2,2)}\del\xi^1\del\xi^2 \del
e^{2i(\varphi^1+\varphi^2)}:(w)\over (z-w)^{1/2}}\ ,
\label{ex1}\eea
where the leading term is given by
 \bea
\S^{(2,1;0)}_{[2];-2}&=&
:c_{-(2,2)}\del\xi^1\del\xi^2e^{2i(\varphi^1+\varphi^2)}:\ .\eea
From the operator product expansions
\bea && E(z)\S^{(2,1;0)}_{[2];-2}(w)\ \sim\
\frac{2}{z-w}\S^{(2,1;0)}_{[2];-2}(w)\
, \qquad M_3(z)\S^{(2,1;0)}_{[2];-2}(w)\ \sim\ 0 \  , \\
&& J_{ij,0}\S^{(2,1;0)}_{[2];-2} \ =\ 0\ , \eea
it follows that $\S_{[2];-2}^{(2,1;0)}$ can be identified with the
$\ell=0$ ground state of the massless vector representation
$\sfD_{[2];-2}(2,1)$ defined in \eq{DP}. The subleading term in
\eq{ex1} is an admixture of $L_{-1}\S_{[2];-2}^{(2,1;0)}$ and $\bar
J^{12}_0\S_{[2];-1}(1,0)$. More generally, the twist fields
$\S^{(\pm m\pm1,m;\ell)}_{[\pm2];-m-1}$ corresponding to the
\emph{massless spin-$m$ ground states} of $\sfD_{[\pm2];-1-m}(\pm
m\pm1,m)$ with minimal conformal weight $h=-1-m$, given in
\eq{massless}, arise as the leading terms in the expansions of
products of arbitrary singleton-valued twist fields (these are
standard composite massless representations of the $\msp(4)$
generated by $M_{AB,0}$; see discussion below \eq{DP}). The result
is
\bea \S^{(\pm
m\pm1,m;\ell)}_{[\pm2];-m-1}&=&V_{\mp(m+\ell+1,m-\ell+1)}\s^{(m;\ell)}_\pm\
, \label{Sigmatwo}\eea
where $\s^{(m;\ell)}_+$ and $\s^{(m;\ell)}_-$ are defined in
\eq{sigmadef1} and \eq{sigmadef2}, respectively. One can verify that
\eq{Sigmatwo} agrees with \eq{massless} and enjoys the properties of
a $(\pm 2)$-twisted primary field, \emph{viz.}
\bea && E(z)\S^{(\pm m\pm1,m;\ell)}_{[\pm2];-m-1}(w)\ \sim\
\frac{\pm(m+1)}{z-w}\S^{(\pm m\pm1,m;\ell)}_{[\pm2];-m-1}(w)\ ,
\\[4pt]
&& M_3(z)\S^{(\pm m\pm1,m;\ell)}_{[\pm2];-m-1}(w)\ \sim\ \frac{\pm
\ell}{z-w}\S^{(\pm m\pm1,m;\ell)}_{[\pm2];-m-1}(w) \ , \\[4pt]
 &&
J_{ij,0}\S^{(m+1,m;\ell)}_{[2];-m-1}(z) \ =\ 0\ , \quad \bar
J^{ij}_{0}\S^{(-m-1,m;\ell)}_{[-2];-m-1}(z) \ =\ 0\ . \eea

The fusions between $P=2$ fields and $P=-2$ fields produce
descendants either to the identity or the quartet representation.
For example we have that
\bea && \S^{(1,0)}_{[2];-1}(z)\S^{(-1,0)}_{[-2];-1}(w)\ \sim\
(z-w)^2 \ ,
\\[4pt]
&& \S^{(1,0)}_{[2];-1}(z)\S^{(-3/2,1/2;1/2)}_{[-2];-3/2}(w)\ \sim\
(z-w)^3 a_1(w)\ , \\[4pt] &&
\S^{(2,1;0)}_{[2];-2}(z)\S^{(-1,0)}_{[-2];-1}(w)\ \sim\ (z-w)^4 \bar
J^{12}(w)\ ,\eea
where the symplectic boson $a_1(z)$ belongs to ${\widehat{\mathfrak
D}}_{[0];1/2}(-1/2,1/2)$. It is also interesting to examine the
fusion between twist fields with $P=2$ and $P=-1$. For instance,
\bea   && \S^{(1,0)}_{[2];-1}(z)\S^{(-1/2,0)}_{[-1];-1/4}(w)\ \sim\
(z-w)\S^{(1/2,0)}_{[1];-1/4}(w) \ , \\[4pt]  &&
\S^{(1,0)}_{[2];-1}(z)\S^{(-1,1/2;1/2)}_{[-1];-1/4}(w)\ \sim\
-2(z-w)^2(J_{11,-1}\S^{(1,1/2;1/2)}_{[1];-1/4})(w) \ ,  \\[4pt] &&
\S^{(2,1;0)}_{[2];-2}(z)\S^{(-5/2,3/2;3/2)}_{[-1];-1/4}(w)\ \sim \
6(z-w)^4(J_{11,-2}\S^{(1,1/2;-1/2)}_{[1];-1/4})(w)\ . \eea
In summary, the free-field realization confirms the fusion rules
\eq{fusprod} derived in Section \ref{Sec:3} using spectral flow.


\scss{On Massive Sector and Realization of Spectral
Flow}\label{Sec:5.5}


In order to demonstrate how the spectral flow operation acts in the
free field basis, let us begin by looking more closely at the twist
fields $\S^{(m\s +P/2,m;\ell)}_{[P];h_{[P],m}}$ with $|P|>2$
($\s={\rm sign} P$) that minimize\footnote{There also exist other
massive fields in the model with higher conformal weights which will
not be considered here. The maximal conformal weight for a given
spin $m$ is given by $h'_{[P],m}=m(|P|-1)-P^2/4$.} the conformal
weight for a given spin $m$, namely
\bea \S^{(m\s
+P/2,m;\ell)}_{[P];h_{[P],m}}&=&\s^{(m;\ell)}_\s\exp\left\{i
\left(\s(m+\ell)+\ft{P}2)\varphi^1+(\s(m-\ell)+\ft{P}2)\varphi^2\right)\right\}
\ ,\\[4pt] h_{[P],m}&=&m(1-|P|)-P^2/4\ ,\eea
where we have suppressed the cocycle factor. These fields
obey\footnote{The detailed fusion rules between generic massive
representations are complicated, although bound to be of the form
\eq{fusprod}, given the validity of composition and distribution
rules in \eq{Gab1} and \eq{Gab}.}
\bea && E(z)\S^{(m\s +P/2,m;\ell)}_{[P];h_{[P],m}}(w)\ \sim\
\frac{m\s +P/2}{z-w}\S^{(m\s +P/2,m;\ell)}_{[P];h_{[P],m}}(w)\ ,
\\[4pt]
&& M_3(z)\S^{(m\s +P/2,m;\ell)}_{[P];h_{[P],m}}(w)\ \sim\ \frac{\pm
\ell}{z-w}\S^{(m\s +P/2,m;\ell)}_{[P];h_{[P],m}}(w) \ , \\[4pt]
\la{lwc1} && J_{ij,0}\S^{(m+P/2,m;\ell)}_{[P];h_{[P],m}}(z) \ =\ 0\
,\qquad P>0\ ,\\[4pt] && \bar
J^{ij}_{0}\S^{(-m+P/2,m;\ell)}_{[P];h_{[P],m}}(z) \ =\ 0\ , \qquad
P<0\ ,\eea
corresponding to \emph{massive ground states} in
$\sfD_{[P];h_{[P],m}}(m+P/2,m)\subset \sfD_{[P]}$. We note that the
special twist fields $\S^{(P/2,0)}_{[P];h_{[P],0}}$, which realize
$P$-twisted primary fields in the sense of \eq{twpr1} and
\eq{twpr2}, obey conditions that are stronger than those in
\eq{lwc1}. We next expand the twist fields in modes,
\bea \S^{(e,m;\ell)}_{[P];h_{[P],m}}(z) &=&
\sum_{l\in\integ-h_{[P],m}}
\left(\S^{(e,m;\ell)}_{[P];h_{[P],m}}\right)_l~z^{-l-h_{[P],m}}\
,\eea
where $e=m\s+P/2$. The corresponding states,
\bea
|[P];h_{[P],m};e,m;\ell\rangle&=&\lim_{z\rightarrow0}\S^{(e,m;\ell)}_{[P];h_{[P],m}}(z)|0\rangle\;=\;
\left(\S^{(e,m;\ell)}_{[P];h_{[P],m}}\right)_{-h_{[P],m}}|0\rangle\
, \eea
thus describe singletons for $|P|=1$, and massless as well as
massive ground states with minimal conformal weight for $|P|>1$.
Inserting the mode expansions of the free fields yields (dropping
a constant multiplicative factor)
\bea \la{mgss} |[P];h_{[P],m};e,m;\ell\rangle\ =\
\chi_\s^{(m;\ell)}e^{ i[\s(m+\ell)+P/2]q^1+
i[\s(m-\ell)+P/2]q^2}|0\rangle\ , \eea
where
\bea \la{chidef}\chi_+^{(m;\ell)}\ =\
\xi^1_{-1}\cdots\xi^1_{-m-\ell}\xi^2_{-1}\cdots\xi^2_{\ell-m} \ ,
\quad \chi_-^{(m;\ell)}\ =\
\eta^1_{-1}\cdots\eta^1_{1-m-\ell}\eta^2_{-1}\cdots\eta^2_{1+\ell-m}\
.\eea
As expected, the $\a^i_{n}$ oscillators drop out in the sector with
minimal conformal weight. We also note that the free-field momenta
$(p^1,p^2)=-(\s(m+\ell)+P/2,\s(m-\ell)+P/2)$ are related to energy
eigenvalues and spin projections by the simple formulas
$e=-(p^1+p^2)/2$ and $\ell=-(p^1-p^2)/2$. The twisted primary ground
states $|[P]\rangle$ in \eq{agsofws} assume the particularly simple
form\footnote{The dual states $\langle0|=(|0\rangle)^\dagger$ obey
$\langle0|0\rangle=1$, which induces the inner product
$\langle[P']|[P]\rangle=\d_{P+P',0}$, consistent with the definition
in \eq{normsP}. The states given in \eq{mgss} have dual
representations given by
$\langle[P];h_{[P],m};e,m;\ell|=(|[P];h_{[P],m};e,m;\ell\rangle)^\dagger$.}
\bea |[P]\rangle&=&e^{\ft{iP}2(q^1+q^2)}|0\rangle \ , \eea
with momentum $(p^1,p^2)=(-P/2,-P/2)$ and conformal weight
$h=-p^ip^i/2=-P^2/4$. These states obey the conditions in
\eq{xx1}--\eq{xx4} by construction, with the affine generators
realized in terms of the $\a^i_{n}$ and $(\xi^i_n,\eta^i_{n})$
oscillators. Thus, the spectral flow operation $\O_P$ acts on
states (kets) by multiplication with $\exp(iP(q^1+q^2)/2)$. This
operation changes the energy eigenvalues and the conformal weights
but clearly does not change the spin.


\scs{Conclusions and Outlook}\label{sec:6}


We have examined the $\widehat{\mso}(2,D-1)$ WZW model at the
subcritical level $k=-(D-3)/2$. It has a singular vector, given
by \eq{KMCR1}, at Virasoro level $2$ in the NS sector whose
decoupling induces a spectrum of KM twisted primary scalars
$\Sigma_{[P]}$, $P\in \integ$, connected to the identity by $P$
units of spectral flow and forming an operator algebra with fusion
rule $\Sigma_{[P]}\times \Sigma_{[P']}=\Sigma_{[P+P']}$. The
decoupling, or the hyperlight-likeness condition \eq{decoupl},
constitutes an affine extension of the equation of motion of the
$(D+1)$-dimensional conformal particle, \emph{i.e.}~the scalar
singleton. For $P\neq 0$, the KM module built on $\Sigma_{[P]}$
contains a unitary subspace of representations in the tensor product
of $|P|$ singletons or anti-singletons for $P>0$ and $P<0$,
respectively. In the special case $D=4$, we have shown that the
spinor singleton and its composites also solve the decoupling
condition \eq{decoupl}. Moreover, in this case, by exploiting the
isomorphism $\mso(2,3)\sim \msp(4)$, we have considered the
$\widehat{\msp}(4)_{-1/2}$ WZW model admitting a realization in
terms of 4 real-symplectic bosons, containing both scalar and spinor
singletons together with their composites. A bosonization procedure
leading to a free-field model has allowed us in particular to
compute the fusion rules explicitly for $\vert P\vert \le 2$ and
compare with the predictions obtained by spectral flow arguments. We
indeed find an agreement. These results provide an embedding of the
Flato-Fr\o nsdal compositeness theorem, stating that massless fields
are made up from singletons in AdS, in a conformal field theory
setting.

The massless sector $P=2$ is particularly interesting for the
purpose of making contact with higher-spin gauge theory. As noted
below \eq{fus1}, the product between a singleton twist field and an
anti-singleton twist field generates an element in the space ${\cal
A}=Env(\msp(4))/{\cal I}$, where ${\cal I}$ is the ideal generated
by the singular vacuum vector $V_{AB}$ defined in \eq{VAB1}. The
space ${\cal A}$, which is an associative algebra, plays an
important role in higher-spin gauge theory. In fact, higher-spin
master fields are differential forms taking values in various
subspaces of ${\cal A}$. In particular, the elements in ${\cal A}$
that are odd under the anti-automorphism $\tau$ define the
higher-spin algebra $\mhs(4)\simeq\mhso_0(4,2)$. It would be
interesting to spell out the exact relation between the higher-spin
algebra and the affine algebra.

Another interesting observation deserving further investigation is
related to the modular invariance of the theory and to the locality
of the operator algebra. In fact, since $\Sigma_{[P]}$ are scalar
particles, the corresponding vertex operators should have a local
operator product with Grassmann even statistics, \emph{i.e.} the
monodromy matrix \eq{monodromy1} should be equal to $1$ for
$\l=(P/2,P/2)$ and $\l'=(P'/2,P'/2)$. This would imply that $P\in
2\integ$. More generally, excited states in the $P$-twisted sector,
created by $N$ symplectic bosons, carry tensorial representations of
$Sp(4)$ for $N$ even and spinorial representations for $N$ odd. The
locality properties of the operator product correspond to
appropriate Grassmann statistics for $P,N\in 2\integ$.
Interestingly, for these values also the conformal weight becomes an
integer, which assures invariance under the modular transformation
$T$ without the need to include an anti-holomorphic sector. It would
be interesting to perform a complete analysis of the locality of the
vertex operator algebra.

We would also like to comment on similar realizations in terms of
symplectic spinor bosons in $D=5$ and $D=7$. We expect these models
to provide affine (massive) extensions of the massless models
constructed in \cite{Sezgin:2001zs,Sezgin:2001ij}.  Here,
degeneracies among the states are lifted by internal gauge
symmetries, based on $U(1)$ and $SU(2)$, respectively, in $D=5$ and
$D=7$. In fact, as shown in \cite{Engquist:2005yt}, the affine
extension is critical in $D=7$ (where it is related by triality to a
critical $Sp(2)$ gauged model based on symplectic vector bosons).
Studying these explicit realizations would be particularly
interesting since, in principle, for special values of $D$ extra
singular vacuum vectors could appear, pointing to striking
differences among models with different values of $D$. An example of
this behavior in $D=7$ will be discussed in a forthcoming paper
\cite{wip}.

In \cite{wip}, we will be mainly concerned with the gaugings of the
$\widehat{\mso}(2,D-1)_{-\e_0}$ WZW model. In fact, we expect that
the gauging of a proper subalgebra $\mh\subset\mg$, such as that in
\eq{compactg}, would remove all (or almost all) of the non-zero mode
excitations in the spectrum. We would then be left with a
topological model containing in its spectrum scalar (and, in
$D=4$, spinor) (anti)singletons as well as all of their composites.
The gauging, however, would not necessarily remove all non-unitary
states. In particular, in $D=4$ the non-unitary anti-spinor
singleton zero-mode representation $\sfD^-_{1/2}$ would survive. The
truncation of this representation, along with other representations
with the wrong space-time statistics, should be controlled by a
GSO-like projection. We expect the resulting GSO-projected gauged
WZW model to be unitary and to consist of scalar singleton and
anti-singleton representations together with all of their tensor
products.

Finally, we would like to say a few words on how a proper gauging of
the WZW models discussed in this paper could fit in a bigger picture
hopefully providing an alternative approach to string quantization
in AdS spacetime. As mentioned in the Introduction, an alternative
to the conventional interpretation of the target space as a
space-time manifold is to regard it as an internal fiber
\cite{Engquist:2005yt}. The WZW model would then be thought of as a
device to realize the internal symmetry algebra, together with a
star product and a trace operation. The spacetime would appear only
upon the operation of unfolding \cite{Bekaert:2005vh}. A crucial
test of this idea would then be to set the issue whether unfolding
could be implemented consistently in conjunction with the WZW model
and, if so, whether the correct free space-time field equations
\cite{Fronsdal:1978vb} for the fields in the model could be
reproduced. Our work is a at very preliminary stage and the
understanding of these issues is left for future work.

{\bf Acknowledgements:}
  This work is supported in part by the European Community's Human Potential Programme under contracts MRTN-CT-2004-005104 {\it Constituents, fundamental forces and symmetries of the universe}, MRTN-CT-2004-503369 {\it The Quest For Unification: Theory Confronts Experiment} and MRTN-CT-2004-512194 {\it Superstring Theory};  and by the INTAS contract
03-51-6346 {\it Strings, branes and higher-spin fields}.
  L.T. would like to thank the Foundation Boncompagni Ludovisi, n\'ee Bildt, for financial support during her two-year stay in Uppsala, where part of the results presented in this paper were obtained. Finally, the research of P.S. is supported in part by a
visiting professorship issued by Scuola Normale Superiore; by INFN;
by the MIUR-PRIN contract 2003-023852; and by the NATO grant PST.CLG.978785.

\begin{appendix}


\scs{Decomposition Formulae for $P=0,1$}\la{sec:appA}


In this appendix we shall derive the decomposition \eq{FPdec} of
the $P$-twisted oscillator Fock spaces $\widehat {\cal F}_{[P]}$
into irreps $\widehat \sfD_{[P]}(e_0,s_0)$ of
$\widehat\mso(2,3)_{-1/2}$. The decompositions for different
values of $P$ are related by spectral flow, which means that it
suffices to show \eq{FPdec} for one value of $P$. In this appendix
we shall do this for $P=1$ and $P=0$ using oscillator methods
(without resorting to spectral flow), viewing one of the cases as
a check of the spectral flow formalism. In general, the Fock
spaces factorize
\bea \widehat {\cal F}_{[P]}&=& \widehat {\cal
F}_{[P]}^{(1)}\otimes \widehat {\cal F}_{[P]}^{(2)}\
,\label{factorP}\eea
where $\widehat {\cal F}_{[P]}^{(i)}$ generated by the action of
$a^i(z)$ and $a_i(z)$ on $\vert[P]\rangle$ for fixed $i=1,2$. The
$\widehat\mso(2,3)_{-1/2}$ irreps decompose in a corresponding way
under $\widehat\mso(2,3)_{-1/2}\rightarrow
\widehat\msp(2)^{(1)}_{-1/2}\oplus\widehat\msp(2)^{(2)}_{-1/2}$,
where the affine $sp(2)$ currents are given in \eq{sp2curr}. Thus,
in order to decompose $\widehat {\cal F}_{[P]}$ under
$\widehat\mso(2,3)_{-1/2}$ we can first decompose $\widehat {\cal
F}_{[P]}^{(i)}$ under $\widehat\msp(2)^{(i)}_{-1/2}$ and then
examine the action of the off-diagonal $\widehat\mso(2,3)_{-1/2}$
currents on the tensor product.

\begin{center}{\it Case $P=1$:}\end{center}
\underline{\textit{Lemma 1:}} For fixed $i$, the Fock space
$\widehat {\cal F}_{[1]}^{(i)}$ decomposes under
$\widehat\msp(2)^{(i)}_{-1/2}$ as
\bea \widehat {\cal F}_{[1]}^{(i)}&=& \widehat
\sfD^{(i)}_{[1]}(\ft14)\oplus \widehat\sfD^{(i)}_{1]}(\ft34)\
,\eea
where $\widehat\sfD^{(i)}_{[1]}(j)$ are built on the two
metaplectic representations $\sfD(j)$ of $\msp(2)$ with
$j=1/4,3/4$,
\bea J^{+}_n|j\rangle^{(i)}&=& 0\quad \mbox{for $n\geq1$}\ ,\\
(J^3_n-j\delta_{n0})|j\rangle^{(i)}&=&0\quad \mbox{for $n\geq 0$}\
,\\J^-_{n}|j\rangle^{(i)}&=&0\quad\mbox{for $n\geq -1$}\
.\label{A5}\eea
We note that $\sfD(1/4)$ and $\sfD(3/4)$ are isomorphic to the
even and odd states, respectively, of the Fock space of a single
oscillator (in \eq{A5}, the $J^-_{-1}$ condition removes a
singular vector that is identically zero in the oscillator
realization).

\underline{\textit{Lemma 2:}} The off-diagonal
$\widehat\msp(4)_{-1/2}$ charges $\{\bar
J^{12}_m,K^{12}_n,K^{21}_n, J^{12}_n\}$ act on tensor products as
follows:
\bea \widehat \sfD_{[1]}^{(1)}(j)\otimes \widehat
\sfD_{[1]}^{(2)}(j')&\mapsto& \widehat
\sfD_{[1]}^{(1)}(1-j)\otimes \widehat \sfD_{[1]}^{(2)}(1-j')\
.\eea
It follows that $\widehat\msp(4)_{-1/2}$ acts irreducibly on
\bea \widehat \sfD_{[1]}(\ft12,0)&=&
\left[\widehat\sfD_{[1]}^{(1)}(\ft14)\otimes
\widehat\sfD_{[1]}^{(2)}(\ft14)\right]\oplus
\left[\widehat\sfD_{[1]}^{(1)}(\ft34)\otimes
\widehat\sfD_{[1]}^{(2)}(\ft34)\right]\ ,\label{irrep1}\\[4pt]
\widehat \sfD_{[1]}(1,\ft12)&=&
\left[\widehat\sfD_{[1]}^{(1)}(\ft14)\otimes
\widehat\sfD_{[1]}^{(2)}(\ft34)\right]\oplus
\left[\widehat\sfD_{[1]}^{(1)}(\ft34)\otimes
\widehat\sfD_{[1]}^{(2)}(\ft14)\right]\ .\label{irrep2}\eea
We note that the ground state of $\widehat\sfD_{[1]}(\ft12,0)$ is
given by $|1/4\rangle^{(1)}\otimes |1/4\rangle^{(2)}$, while
$|3/4\rangle^{(1)}\otimes |3/4\rangle^{(2)}=\bar
J^{12}_0|1/4\rangle^{(1)}\otimes |1/4\rangle^{(2)}$.

Combining Lemmas 1 and 2 with the factoring formula \eq{factorP}
we conclude that $\widehat {\cal F}_{[1]}$ decomposes into the
direct sum of \eq{irrep1} and \eq{irrep2}, as stated in \eq{FPdec}
for $P=1$.

\textit{Derivation of Lemma 1:} To simplify the notation, let us
drop the $(i)$ superscripts and define $(\bar a_n,a_n)=(\bar
a^i_n,a_{i,n})$ and $\widehat{\cal F}=\widehat{\cal
F}_{[1]}^{(i)}$. We decompose $\widehat{\cal F}$ into Virasoro
levels,
\bea \widehat{\cal F}&=& \bigoplus_{\ell=0}^\infty\widehat{\cal
F}_{\ell}\ ,\qquad (L_0-\ell+\ft18)\widehat{\cal F}_{\ell}\ =\ 0\
.\eea
We seek the decomposition into irreducible $\widehat
\msp(2)_{-1/2}$ representations,
\bea  \widehat{\cal F}&=& \bigoplus_{\ell,j}\widehat
\sfD_{\ell}(j)\ ,\eea
where $\widehat \sfD_{\ell}(j)$ is built on the ground state
$|\ell;j\rangle$ at level $\ell$ with $\msp(2)$ spin $j$. There
are two ground states at level $0$, namely
\bea |0;\ft14\rangle&=& |[1]\rangle\ ,\qquad |0;\ft34\rangle\ =\
\bar a_0|[1]\rangle\ .\eea
Let us define
\bea \widehat{\cal M}&=&\widehat \sfD_{0}(\ft14)\oplus \widehat
\sfD_{0}(\ft34)\ =\ \bigoplus_{\ell}\widehat{\cal M}_\ell\ ,\eea
and consider the spaces $\widehat {\cal Q}_{\ell}=\widehat{\cal
F}_\ell/\widehat{\cal M}_{\ell}$. By definition, $\widehat
\sfD_{\ell}(j)$ does not contain any singular vectors, so that if
$|\ell\rangle$ is a ground state at level $\ell\geq 1$, then
$|\ell\rangle\in \widehat {\cal Q}_{\ell}$. Thus, if $\dim\widehat
{\cal Q}_{\ell}=0$ for $\ell\geq 1$ then there are no ground
states for $\ell\geq 1$. Conversely, if there are no ground states
for $\ell\geq 1$ then $\dim\widehat {\cal Q}_{\ell}=0$ for
$\ell\geq 1$ (since the first occurrence of $\dim\widehat {\cal
Q}_{\ell}>0$ would be tied to the existence of such a ground
state). Hence, $\dim\widehat {\cal Q}_{\ell}=0$ for $\ell\geq 1$
is equivalent to that there are no ground states for $\ell\geq 1$.
This can be checked explicitly for $\ell\leq 3$ (either by showing
there are no ground states or by simply rearranging oscillator
excitations into the form of KM descendants in $\widehat{\cal
M}$). We proceed by induction. A state $|\ell;N\rangle \in
\widehat{\cal F}_{\ell}$ with fixed affine $\msp(2)$ spin, say
$(J^3_{0}-\ft{N}2-\ft14)|\ell;N\rangle=0$ ($N\in \integ$), can be
expanded as
\bea |\ell;N\rangle&=& \sum_{\tiny\ba{c}\{m\},\{n\}\\\sum m+\sum n
=\ell\\N'\equiv N-\sum_m 1+\sum_n 1\geq
0\ea}A_{\{m\},\{n\}}\prod_{m} \bar a_{-m} \prod_{n}
a_{-n}|N'\rangle\ ,\eea
where $A_{\{m\},\{n\}}$ are constants, $\{m\}$ and $\{n\}$ are
sets of positive integers, and $|N'\rangle=(a_0)^{N'}|[1]\rangle$
belong to $\sfD(1/4)$ and $\sfD(3/4)$ for $N'$ even and odd,
respectively. In each monomial, at least one $m$ or $n$ must be
positive. Suppose $m>0$. By the induction assumption, the monomial
can then be rewritten as $\bar a_{-m}\sum_{\{k,\a\}}\prod_{k,\a}
J^\a_{-k}|N''\rangle$ where $k$ are positive integers and $N''$ is
fixed by spin conservation. By moving the KM charges to the left,
the monomial can be rearranged into $a_{-\ell}$ and $\bar
a_{-\ell}$ excitations plus descendants in $\widehat{\cal
M}_{\ell}$. An analogous statement holds if $n>0$. Thus,
\bea |\ell;N\rangle&=& A \bar
a_{-\ell}|N-1\rangle+Ba_{-\ell}|N+1\rangle+|\ell;N;{\rm
desc}\rangle\ ,\qquad |\ell;N;{\rm desc}\rangle\in \widehat{\cal
M}_{\ell}\ ,\qquad\eea
where $A$ and $B$ are constants (we note that this shows that
$\dim \widehat {\cal Q}_{\ell}\leq 2$), so that $\dim\widehat
{\cal Q}_{\ell}=0$ if
\bea J^\a_{1}|\ell;N\rangle&=&0\ \Rightarrow\ A\ =\ B\ =\ 0\ .\eea
We note that the absence of singular vectors in $M_{\ell}$ assures
that for fixed $A$ and $B$ the constraint
$J^\a_{1}|\ell;N\rangle=0$ has a unique solution for $|N,\ell;{\rm
desc}\rangle$ (which may of course be trivial), since if
$|N,\ell;{\rm desc}\rangle'$ is another solution then
$J^\a_{1}(|N,\ell;{\rm desc}\rangle-|N,\ell;{\rm desc}\rangle')=0$
implies $|N,\ell;{\rm desc}\rangle-|N,\ell;{\rm desc}\rangle'=0$.
We now expand
\bea |N,\ell;{\rm desc}\rangle&=&\sum_{k=2}^\ell
|N,\ell;(k)\rangle\ ,\eea
where $|N,\ell;(k)\rangle$ is $k$'th order in oscillators
$(a_{-n},\bar a_{-n})$ with $n\geq 1$. The strategy is to work
order by order in $k$ and arrive at some finite order at $A=B=0$
as a compatibility condition. Canceling the linear terms in the
$J^{+}_1$ and $J^-_{1}$ conditions yields
\bea |N,\ell;(2)\rangle&=&-B\bar
a_{-\ell+1}a_{-1}|N\rangle-{A\over N} a_{-\ell+1}\bar
a_{-1}|N\rangle\ .\eea
Canceling the linear terms in the $J^3_1$ condition then yields
the compatibility condition
\bea A+NB&=&0\ .\eea
In the next order one finds
\bea |N,\ell;(3)\rangle&=& {A\over N} \bar a_{-\ell+2}\bar
a_{-1}a_{-1}|N-1\rangle+{B\over N+1}a_{-\ell+2}a_{-1}\bar
a_{-1}|N+1\rangle\ ,\eea
and the compatibility conditions (for $\ell>3$)
\bea A-NB&=&0\ ,\qquad {N-1\over N}A\ =\ 0\ ,\qquad {1\over N+1}B\
=\ 0\ ,\eea
implying $A=B=0$, which completes the proof of Lemma 1.
\begin{center}{\it Case $P=0$:}\end{center}
The analysis parallels that of $P=1$. Let us assume the
decomposition
\bea \widehat {\cal F}_{[0]}^{(i)}&=& \widehat
\sfD_{[0]}^{(i)}(0)\oplus \widehat\sfD_{[0]}^{(i)}(-\ft12)\
,\label{decP1}\eea
with $\widehat\msp(2)^{(i)}_{-1/2}$ ground states given by the
singlet $|[0]\rangle$ and the doublet $(a^i_{-1/2}|[0]\rangle,\bar
a^i_{-1/2}|[0]\rangle)$ (fixed $i$), with lowest $\msp(2)^{(i)}$
spins $j=0$ and $j=-1/2$, respectively. The off-diagonal
$\widehat\msp(4)_{-1/2}$ charges then act on tensor products as
follows
\bea \sfD_{[0]}^{(1)}(j)\otimes \sfD_{[0]}^{(2)}(j')&\mapsto&
\sfD_{[0]}^{(1)}(-\ft12-j)\otimes \sfD_{[0]}^{(2)}(-\ft12-j')\
,\eea
implying that $\widehat\msp(4)_{-1/2}$ acts irreducibly on
\bea \widehat \sfD_{[0]}(0,0)&=& \left[\sfD_{[0]}^{(1)}(0)\otimes
\sfD_{[0]}^{(2)}(0)\right]\oplus\left[\sfD_{[0]}^{(1)}(-\ft12)\otimes
\sfD_{[0]}^{(2)}(-\ft12)\right]\ ,\\[4pt]
\widehat \sfD_{[0]}(1,\ft12)&=& \left[\sfD_{[0]}^{(1)}(0)\otimes
\sfD_{[0]}^{(2)}(-\ft12)\right]\oplus
\left[\sfD_{[0]}^{(1)}(-\ft12)\otimes \sfD_{[0]}^{(2)}(0)\right]\
.\eea
Combined with the factoring formula \eq{factorP}, this yields
\eq{FPdec} for $P=0$. Finally, to derive \eq{decP1} we decompose
into Virasoro levels $\ell\in \{0,\ft12,1,\ft32,\dots\}$. The
first two levels contain the ground states, and there are no new
ground states at levels $\ell=1$ and $\ell=3/2$. We proceed by
induction on $\ell$, treating $\ell\in\integ+1/2$ and
$\ell\in\integ$ as separate cases. The steps are similar to those
for $P=1$. For example, in case $\ell\in\integ+1/2$, the induction
assumption implies $|\ell\rangle=(A a_{-\ell}+B \bar
a_{-\ell})|0\rangle+|\ell;{\rm desc}\rangle$ where $|\ell;{\rm
desc}\rangle=\sum_{k\geq 1}|\ell;(2k+1)\rangle\in \widehat
\sfD_{[0]}(-\ft12)$. The $J^\pm_1$ conditions then yield
\bea |\ell;(3)\rangle&=& (-A a_{-\ell+1}(\bar a_{-1/2})^2+B\bar
a_{-\ell+1}(a_{-1/2})^2)|0\rangle\ ,\eea
which is annihilated by $J^3_1$ to lowest order in oscillators
only for $A=B=0$ (for $\ell\geq 3/2$).

\end{appendix}


\end{document}